%% file: main.tex
\documentclass{lmcs}

\keywords{type-based termination, sized types, futures, concurrency, infinite proofs}

\title{Type-Based Termination for Futures}
\titlecomment{Short version appeared in FSCD 2022.}

\author[S. Somayyajula]{Siva Somayyajula\lmcsorcid{0000-0002-9654-8411}}
\author[F. Pfenning]{Frank Pfenning}

\address{Carnegie Mellon University}
\email{\{ssomayya, fp\}@cs.cmu.edu}

\thanks{This material is based upon work supported by the United States Air Force and DARPA
under Contract No. FA8750-18-C-0092.}

\usepackage{stmaryrd}
\usepackage{mathrsfs,mathtools}
\usepackage{bm, bbm}
\usepackage{cancel}
\usepackage{float}
\usepackage{amsmath, amssymb}

\usepackage{dashrule}
\usepackage{proof-dashed}

\input{defs.tex}

\newcommand{\textcolorbf}[2]{#2}

\newif\iflongversion
\longversiontrue

\begin{document}

\begin{abstract}
In sequential functional languages, sized types enable termination checking of programs with complex patterns of recursion in the presence of mixed inductive-coinductive types. In this paper, we adapt sized types and their metatheory to the concurrent setting. We extend the semi-axiomatic sequent calculus, a subsuming paradigm for futures-based functional concurrency, and its underlying operational semantics with recursion and arithmetic refinements. The latter enables a new and highly general sized type scheme we call \emph{sized type refinements}. As a widely applicable technical device, we type recursive programs with infinitely deep typing derivations that unfold all recursive calls. Then, we observe that certain such derivations can be made infinitely wide but finitely deep. The resulting trees serve as the induction target of our termination result, which we develop via a novel logical relations argument.
\end{abstract}

\maketitle

\section{Introduction}
\emph{Note: this is an extended version of an eponymous paper that appeared in FSCD 2022 that includes further examples (Examples \ref{ex:nat-eater}, \ref{ex:qsort}, and \ref{ex:stream-proc}), a more straightforward presentation of the metatheory (Section \ref{sec:semantics}) based on Kripke logical relations \cite{Plotkin1973}, and a representative set of the corresponding proofs (Sections \ref{sec:intermediate} and \ref{sec:semantics}).}
\\

Adding (co)inductive types and terminating recursion (including productive corecursive definitions) to any programming language is a non-trivial task, since only certain recursive programs constitute valid applications of (co)induction principles. Briefly, inductive calls must occur on data smaller than the input and, dually, coinductive calls must be guarded by further codata output. In either case, we are concerned with the decrease of (co)data size---height of data and observable depth of codata---in a sequence of recursive calls. Since inferring this exactly is intractable, languages like Agda (before version 2.4) \cite{Abel2016jfp} and Coq \cite{Coq} resort to conservative syntactic criteria like the \emph{guardedness check}.

One solution that avoids syntactic checks is to track the flow of (co)data size at the type level with \emph{sized types}, as pioneered by Hughes et al.~\cite{Hughes96popl} and further developed by others \cite{Barthe2004fgpu,Blanqui,Abel2008lmcs,Abel2016jfp}. Inductive and coinductive types are indexed by the height and observable depth of their data and codata, respectively. \textcolorbf{red}{Consider the equirecursive type definitions in Example \ref{ex:recur-ty} adorned with our novel \emph{sized type refinements}}: $\nat[i]$ describes unary natural numbers less than or equal to $i$ and $\str_A[i]$ describes infinite $A$-streams that allow the first $i+1$ elements to be observed before reaching potentially undefined or divergent behavior. $\oplus$ and $\&$ are respectively analogous to eager variant record and lazy record types in the functional setting.

\begin{exa}[Recursive types]
\label{ex:recur-ty}
\begin{align*}
\nat[i]&=\oplus\{\zero:\one,\succc:i>0\land\nat[i-1]\}\\
\str_A[i]&=\&\{\head:A,\tail:i>0\implies\str_A[i-1]\}
\end{align*}
Note that $\str_A[i]$ is \emph{not} polymorphic, but is parametric in the choice of $A$ for demonstrative purposes.
\end{exa}

The phrases $\phi\land\ldots$ and $\phi\implies\ldots$ are \emph{constrained types}, so that the $\succc$ branch of $\nat[i]$ produces a $\nat$ at height $i-1$ \emph{when} $i>0$ whereas the $\tail$ branch of $\str_A[i]$ can produce the remainder of the stream at depth $i-1$ \emph{assuming} $i>0$. Starting from $\nat[i]$, recursing \emph{on}, for example, $\nat[i-1]$ ($i>0$ is assumed during \emph{elimination} so that $i-1$ is well-defined) produces the size sequence $i>i-1>i-2>\ldots$ that eventually terminates at $0$, agreeing with the (strong) induction principle for natural numbers. Dually, starting from $\str_A[i]$, recursing \emph{into} $\str_A[i-1]$ (again, $i>0$ is assumed during \emph{introduction} so that $i-1$ is well-defined) produces the same well-founded sequence of sizes, agreeing with the coinduction principle for streams. In either case, a recursive program terminates if its call graph generates a well-founded sequence of sizes in each code path. Most importantly, the behavior of constraint conjunction and implication during elimination and introduction encodes induction and coinduction, respectively. To see how sizes are utilized in the definition of recursive programs, consider the type signatures below. We will define the code of these programs in Example \ref{ex:odd-even}.

\begin{exa}[Evens and odds I]
\label{ex:odds-evens-ty}
Postponing the details of our typing judgment for the moment, the signature below describes definitions that project the even- and odd-indexed substreams \textcolorbf{blue}{(referred to by $y$)} of some input stream \textcolorbf{blue}{(referred to by $x$)} at half of the original depth. Note that indexing begins at zero.
\begin{align*}
i;\cdot;x:\str_A[2i]&\vdash^i y\from\evens\,i~x::(y:\str_A[i])\\
i;\cdot;x:\str_A[2i+1]&\vdash^i y\from\odds\,i~x::(y:\str_A[i])
\end{align*}
An alternate typing scheme that hides the exact size change is shown below---given a stream of \emph{arbitrary} depth, we may project its even- and odd-indexed substreams of arbitrary depth, too. We provide implementations for both versions in Example \ref{ex:odd-even}.
\begin{align*}
i;\cdot;x:\forall j.\,\str_A[j]&\vdash^i y\from\evens\,i~x::(y:\str_A[i])\\
i;\cdot;x:\forall j.\,\str_A[j]&\vdash^i y\from\odds\,i~x::(y:\str_A[i])
\end{align*}
$\exists j.\,X[j]$ and $\forall j.\,X[j]$ denote \emph{full} inductive and coinductive types, respectively, classifying (co)data of arbitrary size. In general, less specific type signatures are necessary when the exact size change is difficult to express at the type level \cite{Xi99popl}. For example, in relation to an input list of height $i$, the height $j$ of the output list from a list filtering function may be constrained as $j\leq i$.
\end{exa}

Sized types are \emph{compositional}: since termination checking is reduced to an instance of typechecking, we avoid the brittleness of syntactic termination checking. \textcolorbf{red}{However, we find that \emph{ad hoc} features for implementing size arithmetic in the prior work can be subsumed by more general \emph{arithmetic refinements}~\cite{Das20concur,Xi99popl}, giving rise to our notion of sized type refinements that combine the ``good parts'' of modern sized type systems. First, the instances of constraint conjunction and implication to encode inductive and coinductive types, respectively, in our system are similar to the bounded quantifiers in MiniAgda \cite{Abel2012fics}, which gave an elegant foundation for mixed inductive-coinductive functional programming, avoiding continuity checking \cite{Abel2008lmcs}. Unlike the prior work, however, we are able to modulate the specificity of type signatures: (slight variations of) those in Example \ref{ex:odds-evens-ty} are given in $\textsf{CIC}\widehat{\phantom{}_\ell}$ \cite{Sacchini14} and MiniAgda \cite{Abel2012fics,StreamWorkshop}. Furthermore, we avoid transfinite indices in favor of permitting some unbounded quantification (following Vezzosi \cite{Vezzosi2015types}), achieving the effect of somewhat complicated infinite sizes without leaving finite arithmetic.}

\textcolorbf{red}{Moreover, some prior work, which is based on sequential functional languages, encodes recursion via various fixed point combinators that make both mixed inductive-coinductive programming \cite{Basold2018} and substructural typing difficult, the latter requiring the use of the ! modality \cite{Wadler12jfp}. Thus, like $\textsf{F}_{\omega}^{\textsf{cop}}$ \cite{Abel2016jfp}, we consider a signature of parametric recursive definitions. However, we make typing derivations for recursive programs infinitely deep by unfolding recursive calls \emph{ad infinitum} \cite{Brotherston,Lepigre2019}, which is not only more elegant than finitary typing, but also simplifies our termination argument.} To prove termination of program reduction, we observe that \emph{arithmetically closed} typing derivations, which have no free arithmetic variables or constraint assumptions, can be translated to infinitely wide but finitely deep trees of a different judgment. The resulting derivations are then the induction target for our proof, leaving the option of making the original typing judgment arbitrarily rich. Thus, although our proposed language is not substructural, this result extends to programs that use their data substructurally. {In short, our contributions are as follows:}
\begin{enumerate}
\item A general system of sized types based on arithmetic refinements subsuming features of prior systems, such as the mixed inductive-coinductive types of MiniAgda \cite{Abel2012fics} as well as the linear size arithmetic of $\textsf{CIC}\widehat{\phantom{}_\ell}$ \cite{Sacchini14}. Moreover, we do not depend on transfinite arithmetic.
\item The first language for mixed inductive-coinductive programming that is a subsuming paradigm \cite{Levy04} for futures-based functional concurrency.
\item A method for proving termination in the presence of infinitely deep typing derivations by translation to infinitely wide but finitely deep trees.
\end{enumerate}
We define $\text{SAX}^\infty$, which extends the semi-axiomatic sequent calculus (SAX) \cite{DeYoung2020fscd} with arithmetic refinements, recursion, and infinitely deep typing derivations (Section \ref{sec:language}). Then, we define an auxiliary type system called $\text{SAX}^\omega$ which has infinitely wide but finitely deep derivations to which we translate the derivations of $\text{SAX}^\infty$ (Section \ref{sec:intermediate}). Then, we show that all $\text{SAX}^\omega$-typed programs terminate by a novel logical relations argument over \emph{configurations} of processes that capture the state of a concurrent computation (Section \ref{sec:semantics}).

\section{$\text{SAX}^\infty$}\label{sec:language}
In this section, we extend SAX \cite{DeYoung2020fscd} with recursion and arithmetic refinements in the style of Das and Pfenning \cite{Das20concur}. SAX is a logic-based formalism and subsuming paradigm \cite{Levy04} for concurrent functional programming that conceives call-by-need and call-by-value strategies as particular concurrent schedules \cite{Pruiksma20arxiv}. Concurrency and parallelism devices like fork/join, futures \cite{Halstead85}, and SILL-style \cite{Toninho13} monadic concurrency can all be encoded and used side-by-side in SAX \cite{Pruiksma20arxiv}.

To review SAX, let us make observations about proof-theoretic \emph{polarity}. In the sequent calculus, inference rules are either \emph{invertible}---can be applied at any point in the proof search process, like the right rule for implication---or \emph{noninvertible}, which can only be applied when the sequent ``contains enough information,'' like the right rules for disjunction. Connectives that have noninvertible right rules are \emph{positive} and those that have noninvertible left rules are \emph{negative}. The key innovation of SAX is to replace the noninvertible rules with their axiomatic counterparts in a Hilbert-style system. Consider the following right rule for implication as well as the original left rule in the middle that is replaced with its axiomatic counterpart on the right.

\begin{center}
\small\begin{tabular}{c c c}
$\infer[{\to}{\R}]{\Gamma\vdash A\to B}{\Gamma,A\vdash B}$ & $\cancel{\infer[{\to}{\LL}]{\Gamma,A\to B\vdash C}{\Gamma,A\to B\vdash A & \Gamma,A\to B,B\vdash C}}$ & $\infer[{\to}{\LL}]{\Gamma,A\to B,A\vdash B}{}$
\end{tabular}
\end{center}

Since the axiomatic rules drop the premises of their sequent calculus counterparts, cut elimination corresponds to asynchronous communication just as the standard sequent calculus models synchronous communication \cite{Caires10concur}. In particular, SAX has a \emph{shared memory interpretation}, mirroring the memory-based semantics of \emph{futures} \cite{Halstead85}. A future $x$ of type $A$ either contains an object of type $A$ or is not yet populated. A process reading from $x$ either succeeds immediately or blocks if $x$ is not yet populated. As a result, the sequent becomes the typing judgment (extended with arithmetic refinements in the style of \cite{Das20concur}):
\[\overbrace{i,j,\ldots}^{\displaystyle\V};\overbrace{\phi,\psi,\ldots}^{\displaystyle\CC};\overbrace{x:A,y:B,\ldots}^{\displaystyle\Gamma}\vdash^{\overline{e}} P::(z:C)\]
where the \emph{arithmetic variables} in $\V$ are free in the \emph{constraints} (arithmetic formulas) in $\CC$, the types in $\Gamma$, the \emph{process} $P$, and type $C$; moreover, the \emph{address variables} in $\Gamma$, which are free in $P$, \textcolorbf{blue}{stand for addresses of memory cells representing futures. In particular, $P$ reads from $x,y,\ldots$ (\emph{sources}) and writes to $z$ (a \emph{destination}) according to the protocols specified by $A,B,\ldots$ and $C$, respectively. $z$ is written to exactly once corresponding to the population of a future \cite{Halstead85}.} Lastly, the vector (indicated by the overline) of \emph{arithmetic expressions} $\overline{e}$ will be used to track the sizes encountered at each recursive call as mentioned in the introduction. Now, let us examine the definitions of types and processes. For our purposes, detailed syntaxes for expressions $e$ and formulas $\phi$ are unnecessary.
\begin{defi}[Type]

Types are defined by the following grammar, presupposing some mutually recursive type definitions of the form $X[\overline{i}]=A_X(\overline{i})$. Positive types (in the left column) and negative types (in the right column) are colored red and black, respectively. Recursive type names are colored blue since they take on the polarity of their definientia.
\begin{center}
\hspace{-1em}
\small\begin{tabular}{r l l r l l}
$A,B:=$&\color{red}$\mathbf{1}$ & \color{red}unit&$\mid$&\color{blue}{$X[\overline{e}]$} & \color{blue}equirecursive type\\
$\mid$&\color{red}$A\otimes B$ & \color{red}eager pair & $\mid$&$A\to B$ & function\\
$\mid$&\color{red}$\oplus\{\ell:A_\ell\}_{\ell\in S}$ & \color{red}eager variant record & $\mid$&$\&\{\ell:A_\ell\}_{\ell\in S}$ & lazy record \\
$\mid$&\color{red}{$\phi\land A$} & \color{red}constraint conjunction & $\mid$&{$\phi\implies A$} & constraint implication\\
$\mid$&\color{red}{$\exists i.\,A(i)$} & \color{red}arithmetic dependent pair&$\mid$&{$\forall i.\,A(i)$} & arithmetic dep. function
\end{tabular}
\end{center}
\end{defi}
There are eight kinds of processes: two for the structural rules (identity and cut), one for each combination of type polarity (positive or negative) and rule type (left or right), one for definition calls, and one for unreachable code.
\begin{defi}[Process]
Processes are defined by the following grammar. The superscripts $\R$ and $\W$ indicating reading from or writing to a cell.

\begin{center}
\begin{tabular}{r l l}
$P,Q:=$ & $y^{\W}\from x^{\R}$ & copy contents of $x$ to $y$\\\\
 $\mid$ & $x\from P(x);Q(x)$ & allocate $x$, spawn $P$ to write to $x$ and concurrently\\
& &  proceed as $Q$, which may read from $x$\\\\
$\mid$ & $x^{\W}.V$ & write value $V$ to $x$\\\\
$\mid$ & $\casee x^{\R}\,K$ & read value stored in $x$ and pass it to continuation $K$\\\\
$\mid$ & $\casee x^{\W}\,K$ & write continuation $K$ to $x$\\\\
$\mid$ & $x^{\R}.V$ & read continuation stored in $x$ then pass value $V$ to it\\\\
$\mid$ & {$y\from f~\overline{e}~\overline{x}$} & expands to $P_f(\overline{e},\overline{x},y)$ from a signature of mutually\\
& & recursive definitions of the form {$y\from f~\overline{i}~\overline{x}=P_f(\overline{i},\overline{x},y)$}\\\\
$\mid$ & {$\impossible$} & unreachable code due to inconsistent arithmetic context
\end{tabular}
\end{center}

The first two kinds of processes correspond to the identity and cut rules. Values $V$ and continuations $K$ are specified on a per-type-and-rule basis in the following two tables. Note the address variable $x$ distinguished by each rule.

\begin{center}
\begin{minipage}{0.4\textwidth}
\begin{center}
\begin{tabular}{c | c c}
polarity & right rule & left rule\\
\hline
positive & $x^{\W}.V$ & $\casee x^{\R}\,K$\\
negative & $\casee x^{\W}\,K$ & $x^{\R}.V$
\end{tabular}
\end{center}
\end{minipage}
\qquad
\begin{minipage}{0.4\textwidth}
\begin{center}
\begin{tabular}{c | c | c}
type(s) & value $V$ & continuation $K$\\
\hline
$\mathbf{1}$ & $\pair{}$ & $\pair{}\To P$\\
${\otimes},{\to}$ & $\pair{y,z}$ & $\pair{y,z}\To P(y,z)$\\
${\&},{\oplus}$ & $\ell\,y$ & $\{\ell\,y\To P(y)\}_{\ell\in S}$\\
{${\land},{\implies}$} & {$\pair{\ast,y}$} & {$\pair{\ast,y}\To P(y)$}\\
{${\forall},{\exists}$} &
{$\pair{e,y}$} & {$\pair{i,y}\To P(i,y)$}
\end{tabular}
\end{center}
\end{minipage}
\end{center}
\end{defi}

To borrow terminology from linear logic, the ``multiplicative'' group ($\mathbf{1}$, ${\otimes}$, ${\to}$) is concerned with writing addresses, whereas the ``additive'' group (${\oplus}$, ${\&}$) is concerned with writing labels and their case analysis. \textcolorbf{blue}{Constrained types read and write a placeholder $\ast$ indicating that a constraint is asserted or assumed. However, we will suppress instances of $\ast$ in the example code given, since assumptions and assertions are inferrable in the absence of consecutively alternating constraints (e.g., $\phi\land(\psi\implies A)$). On the other hand, the arithmetic data communicated by quantifiers are visible since inference is difficult in general \cite{Das20ppdp}.} Now that we are acquainted with the process syntax, let us complete Example \ref{ex:odds-evens-ty}.

\begin{exa}[Evens and odds II]
\label{ex:odd-even}
Recall that we are implementing the following signature and $\str_A[i]=\&\{\head:A,\tail:i>0\implies\str_A[i-1]\}$.
\begin{align*}
i;\cdot;x:\str_A[2i]&\vdash^i y\from\evens\,i~x::(y:\str_A[i])\\
i;\cdot;x:\str_A[2i+1]&\vdash^i y\from\odds\,i~x::(y:\str_A[i])
\end{align*}
The even-indexed substream retains the head of the input, but its tail is the odd-indexed substream of the input's tail. The odd-indexed substream, on the other hand, is simply the even-indexed substream of the input's tail. Operationally, the heads and tails of both substreams are computed on demand similar to a lazy record. Unlike their sequential counterparts, however, the recursive calls proceed concurrently due to the nature of cut. \textcolorbf{blue}{Since our examples will keep constraints implicit, we indicate when constraints are assumed or asserted inline for clarity.}
\begin{align*}
y\from\evens\,i~x=\casee y^{\W}\{&\head\,h\To x^{\R}.\head\,h,\\
&\hspace{-1em}\underbrace{\tail y_t}_{i>0~\text{assumed}}\hspace{-1em}\To x_t\from \underbrace{x^{\R}.\tail x_t}_{2i>0~\text{asserted}};\underbrace{y_t\from\odds\,(i-1)~x_t}_{i;i>0\vdash i-1<i~\text{checked}} \}
\end{align*}

\begin{align*}
y\from\odds\,i~x=x_t\from \hspace{-0.75em}\underbrace{x^{\R}.\tail x_t}_{2i+1>0~\text{asserted}}\hspace{-0.75em};y\from\evens\,i~x_t
\end{align*}
By inlining the definition of $\odds$ in $\evens$ and vice versa, both programs terminate according to our criterion from the introduction even though $\odds$ calls $\evens$ with argument $i$. However, we sketch an alternate termination argument for similar such definitions at the end of Section \ref{sec:intermediate}. On the other hand, consider the alternate signature we gave.
\begin{align*}
i;\cdot;x:\forall j.\,\str_A[j]&\vdash^i y\from\evens\,i~x::(y:\str_A[i])\\
i;\cdot;x:\forall j.\,\str_A[j]&\vdash^i y\from\odds\,i~x::(y:\str_A[i])
\end{align*}
First, we define head and tail observations on streams of arbitrary depth. Since they are not recursive, we do not bother tracking the size superscript of the typing judgment, since they can be inlined. Moreover, we take the liberty to nest values (boxed and highlighted yellow), which can be expanded into SAX \cite{Pruiksma20arxiv}.
\begin{align*}
&\cdot;\cdot;x:\forall j.\,\str_A[j]\vdash y\from\head x::(y:A)\\
&y\from\head x=x^{\R}.\fbox{\mathcolorbox{yellow}{\pair{0,\head y}}}\\
&\cdot;\cdot;x:\forall j.\,\str_A[j]\vdash y\from\tail x::(y:\forall j.\,\str_A[j])\\
&y\from\tail x=\casee y^{\W}\,(\pair{j,y'}\To \underbrace{x^{\R}.\fbox{\mathcolorbox{yellow}{\pair{j+1,\tail y'}}}}_{j+1>0~\text{asserted}})
\end{align*}
The implementation of odds and evens follows almost exactly as before with the above observations in place. Note that we use the abbreviation $y\from f~\overline{e}~\overline{x};Q\triangleq y\from(y\from f~\overline{e}~\overline{x});Q$ for convenience.
\begin{align*}
&y\from\evens\,i~x=\casee y^{\W}\{\head\,h\To y\from\head x,\\
&\qquad\qquad\quad\qquad\qquad\qquad\qquad~{\tail y_t}\To x_t\from\tail x;{y_t\from\odds\,(i-1)~x_t} \}\\
&y\from\odds\,i~x=x_t\from\tail x;y\from\evens\,i~x_t
\end{align*}
\end{exa}

\begin{figure}[h!]
\centering
\begin{tabular}{c}
\hspace{-6em}$\ruleid\quad\rulecut$\\\\
\hspace{-6em}$\ruleoneR\quad\ruleoneL$\\\\
\hspace{-6em}$\ruleotimesR\quad\ruleotimesL$\\\\
\hspace{-6em}$\rulearrR\quad\rulearrL$\\\\
\hspace{-6em}$\ruleoplusR\quad\ruleoplusL$\\\\
\hspace{-6em}$\rulewithR\quad\rulewithL$\\\\
\hspace{-6em}$\ruleexistsR\quad\ruleexistsL$\\\\
\hspace{-6em}$\ruleforallR\quad\ruleforallL$\\\\
\hspace{-6em}$\textcolorbf{blue}{\ruleqR}\quad\textcolorbf{blue}{\ruleqL}$\\\\
\hspace{-4em}$\textcolorbf{blue}{\rulebR}\quad\textcolorbf{blue}{\rulebL}$\\\\
\hspace{-4em}\infer{\V;\CC;\Gamma\vdash\mathbf{impossible}::(x:A)}{\V;\CC\vdash\bot}\\\\
\hspace{-4em}$\rulein\quad\rulecall$
\end{tabular}
\caption{$\text{SAX}^\infty$ Typing}
\label{tab:proc-typing}
\end{figure}

Refer to Figure \ref{tab:proc-typing} for the full process typing judgment---we will comment on specific rules when necessary, but section 5 of \cite{DeYoung2020fscd} discusses the propositional rules more closely. In particular, the arithmetic typing rules make use of a \emph{well-formedness judgment} $\V;\CC\vdash e$ and \emph{entailment} $\V;\CC\vdash\phi$. Most importantly, there are two rules for recursive calls; let us reproduce them below.
$$\rulein$$
$$\rulecall$$
Our process typing judgment is itself \emph{mixed inductive-coinductive} \cite{Danielsson09}---we introduce the auxiliary judgment $\V;\CC;\Gamma\vdash_\infty^{\overline{e}}P::(y:A)$ that is \emph{coinductively} generated by the $\infty$ rule (indicated by the double line). Since the premise of the call rule refers to $\V;\CC;\Gamma\vdash_\infty^{\overline{e}}P::(y:A)$, all valid typing derivations are trees whose infinite branches have a call-$\infty$ pair occurring infinitely often, representing the unfolding of a recursive process. At each unfolding, we check that the arithmetic arguments have decreased (from $\overline{e}$ to $\overline{e'}$) lexicographically\footnote{If two vectors have different lengths, then zeroes are appended to the shorter one.} for termination.

\textcolorbf{red}{We conjecture that finite-time typechecking only requires restricting our type system (including equality of equirecursive types \cite{Das2020rast}) to \emph{circular derivations}, which can be represented as finite trees with loops, provided decidable arithmetic (e.g., Presburger). Such a restricted system may be put in correspondence with a finitary system that detects said loops \cite{SprengerDam2003,Brotherston,Dagnino,Pruiksma20arxiv} where arithmetic assertions can be discharged mechanically \cite{Das2020rast}.} In Example \ref{ex:nat-eater} below, we show a hypothetical instance of typechecking.

\begin{exa}[Typechecking]
\label{ex:nat-eater}
The process definition below, whose type signature is $i;\cdot;x:\nat[i]\vdash^i y\from\eat~i~x::(y:\one)$, traverses a unary natural number by induction to produce a unit. Recall $\nat[i]=\oplus\{\zero:\one,\succc:i>0\land\nat[i-1]\}$.
\begin{align*}
y\from\eat~i~x=\casee x^{\R}\,\{&\zero\,z\To y^{\W}\from {z}^{\R},\succc\,z\To y\from\eat~(i-1)~z\}
\end{align*}
Now, let us construct a typing derivation of its body below.
$$\small D=\infer[\oplus\text{L}]{i;\cdot;x:\nat[i]\vdash^i(y:\mathbf{1})}{\infer[\text{id}]{z:\mathbf{1}\vdash^i(y:\mathbf{1})}{}\hspace{-3em} & \infer[{\land}{\LL}]{i;\cdot;z:i>0\land\nat[i-1]\vdash^i(y:\mathbf{1})}{\infer[\text{call}]{i;i>0;z:\nat[i-1]\vdash^i(y:\mathbf{1})}{i;i>0\vdash i-1<i & \infer=[\infty]{i;\cdot;z:\nat[i-1]\vdash_\infty^{i-1}(y:\mathbf{1})}{\infer*{i;\cdot;z:\nat[i-1]\vdash^{i-1}(y:\mathbf{1})}{[(i-1)/i][z/x]D}}}}}$$
For space, we omit the process terms. Of importance is the instance of the call rule for the recursive call to eat: the check $i-1<i$ verifies that the process terminates and the loop $[(i-1)/i][z/x]D$ ``ties the knot'' on the typechecking process. Mutually recursive programs, then, are checked by circular typing derivations that are mutually recursive \emph{in the metatheory}.
\end{exa}

Note that in subsequent examples, in lieu of writing unwieldy typing derivations, we will demarcate when arithmetic constraints are assumed and asserted, as well as how a recursive call is checked. Now, to illustrate compositionality of typechecking, i.e., termination checking without full source code availability, we show below how to develop a terminating program against a library thereof.

\begin{exa}[Na\"ive Quicksort]
\label{ex:qsort}
A na\"ive ``functional quicksort'' can be implemented against the following signature of definitions assuming $\nat=\exists i.\,\nat[i]$ and $\listT[n]=\oplus\{\nil:\mathbf{1},\cons:\nat\otimes(n>0\land\listT[n-1])\}$---the latter classifying natural number lists of length at most $n$.
\begin{align*}
&m,n;\cdot;x:\listT[m],y:\listT[n]\vdash z\from\append\,(m,n)\,(x,y)::(z:\listT[m+n])\\
&k;\cdot;p:\nat,x:\listT[k]\vdash y\from\partition k\,(p,x)::(y:\exists m,n.\,k=m+n\land\listT[m]\otimes\listT[n])
\end{align*}
That is, we assume that we have definitions that (1) append two lists together and (2) partitions one by a pivot. Then, at a high level, quicksort is a size-preserving definition with the input list length as its termination measure. For brevity, we nest patterns (boxed and highlighted yellow), which can be expanded into nested matches \cite{Pruiksma20arxiv}.
\begin{align*}
&k;\cdot;x:\listT[k]\vdash y\from\qsort k\,x::(y:\listT[k])\\
&y\from\qsort k\,x=\\
&\quad\casee x^{\R}\,\{\nil x'\To y^{\W}\from x^{\R},\\
&\qquad\qquad\quad\underbrace{\fbox{\mathcolorbox{yellow}{\cons\,\pair{h,t}}}}_{k>0\text{ assumed}}\To z\from\partition\,(k-1)\,(h,t);\\
&\qquad\qquad\qquad\quad\casee z^{\R}\,\{\underbrace{\fbox{\mathcolorbox{yellow}{\pair{m,{n,{l,r}}}}}}_{k-1=m+n\text{ assumed}}\To\underbrace{l'\from\qsort m\,l}_{m<k\text{ checked}};\underbrace{r'\from\qsort n\,r}_{n<k\text{ checked}};\\
&\qquad\qquad\qquad\qquad\qquad\qquad s\from\underbrace{s^{\W}.\cons\pair{h,r'}}_{n+1>0\text{ asserted}};y\from\append\,(m,n+1)\,(l',s)\}\}
\end{align*}
Note that despite not being structurally recursive, quicksort still passes the termination check since size information flows from partition, to recursive calls to quicksort, then finally to append.
\end{exa}

Now, consider the following two examples that demonstrate a use case of mixed induction-coinduction in concurrency.
\begin{exa}[Left-fair streams]
\label{ex:lfair-def}
Let us define the mixed inductive-coinductive type $\lfair_{A,B}[i,j]$ of \emph{left-fair streams} \cite{Basold2018}: infinite $A$-streams where each element is separated by finitely many elements in $B$. Once again, these types are \emph{not} polymorphic, but are parametric in the choice of $A$ and $B$ for demonstration.
\begin{align*}
\lfair_{A,B}[i,j]&=\oplus\{\now:\&\{\head:A,\tail:\lfair^{\nu}_{A,B}[i,j]\},\later:B\otimes\lfair^{\mu}_{A,B}[i,j]\}\\
\lfair^{\nu}_{A,B}[i,j]&=i>0\implies\exists j'.\,\lfair_{A,B}[i-1,j']\\
\lfair^{\mu}_{A,B}[i,j]&=j>0\land\lfair_{A,B}[i,j-1]
\end{align*}

In particular, $i$ bounds the observation depth of the $A$-stream whereas $j$ bounds the height of the $B$-list in between consecutive $A$ elements. Thus, this type is defined by lexicographic induction on $(i,j)$. First, the provider may offer an element of $A$, in which case the observation depth of the stream decreases from $i$ to $i-1$ (in the coinductive part, $\lfair^{\nu}_{A,B}[i,j]$). As a result, $j$ may be ``reset'' as an arbitrary $j'$. On the other hand, if an element of ``padding'' in $B$ is offered, then the depth $i$ does not change. Rather, the height of the $B$-list decreases from $j$ to $j-1$ (in the inductive part, $\lfair^{\mu}_{A,B}[i,j]$). By using left-fair streams, we can model processes that permit some timeout behavior but are eventually productive, since consecutive elements of type $A$ are interspersed with only finitely many timeout acknowledgements of type $B$. Armed with this type, we can define a \emph{projection} operation \cite{Basold2018} that removes all of a left-fair stream's timeout acknowledgements concurrently, returning an $A$-stream. As before, nested patterns are boxed and highlighted.
\begin{alignat*}{3}
&i,j;\cdot;x:\lfair_{A,B}[i,j]\vdash^{(i,j)} y\from\san\,(i,j)~x::(y:\str_A[i])\\
&y\from\san\,(i,j)~x=\\&\casee x^{\R}\,(\now s\To\casee y^{\W}(\head h\To s^{\R}.\head h,\\
&\qquad\qquad\qquad\qquad\qquad\qquad~~\underbrace{\tail t}_{i>0~\text{assumed}}\hspace{-1em}\To u\from s^{\R}.\tail u;\\
&\qquad\qquad\qquad\qquad\qquad\qquad\qquad\qquad\casee u^{\R}\,(\hspace{-0.8em}\underbrace{\pair{j',x'}}_{i>0~\text{asserted}}\To \hspace{-1.3em}\underbrace{t\from\san\,(i-1,j')~x'}_{i,j,j';i>0\vdash(i-1,j')<(i,j)~\text{checked}}\hspace{-1em})),\\
&\qquad\qquad\underbrace{\fbox{\mathcolorbox{yellow}{\later\pair{b,x'}}}}_{j>0~\text{assumed}}\To\underbrace{y\from\san\,(i,j-1)~x'}_{i,j;j>0\vdash(i,j-1)<(i,j)~\text{checked}})
\end{alignat*}
%Essentially, we want to convert a $\chan_A[i,j]$ to a $\str_A[i]$ by lexicographic induction on $(i,j)$. In case the head element is produced, it is copied to the output stream. Concurrently, the rest of the channel is recursively sanitized and outputted---the input depth decreases to $i-1$, allowing $j$ to stay the same. Otherwise, if padding is encountered, it is dropped and the channel is sanitized recursively at the lower height $j-1$ ($i$ stays the same).
\end{exa}

\begin{exa}[Stream processors]
\label{ex:stream-proc}
The mixed inductive-coinductive type $\spp_{A,B}[i,j]$ of stream processors of input depth $i$ and output depth $j$ represents \emph{continuous} (in the sense of \cite{Ghani09}) functions from $\str_A[i]$ to $\str_B[j]$; we define it below. Ghani et al.~\cite{Ghani09} define it as the nested greatest-then-least fixed point $\nu X.\,\mu Y.\,(A\times Y)+(B\times X)$, but we adapt the version by Abel \cite{Abel2012fics} to finite size arithmetic.

\begin{align*}
\spp_{A,B}[i,j]&=\oplus\{\get:A\to\spp^{\mu}_{A,B}[i,j],\pput:\&\{\now:B,\rest:\spp^{\nu}_{A,B}[i,j]\}\}\\
\spp^{\mu}_{A,B}[i,j]&=i>0\land\forall j'.\,\spp_{A,B}[i-1,j']\\
\spp^{\nu}_{A,B}[i,j]&=j>0\implies\spp_{A,B}[i,j-1]
\end{align*}

Such functions may consume finitely many elements of type \(A\) from the input stream (the inductive part $\spp^\mu_{A,B}[i]$) before outputting arbitrarily many elements of type \(B\) onto the output stream (the coinductive part $\spp^\nu_{A,B}[j]$). This requires a lexicographic induction on \((i,j)\)---in the inductive part, the input depth decreases to $i-1$, so the new output depth $j'$ may be arbitrary. In the coinductive part, $i$ stays the same, so $j$ must decrease (to $j-1$). Let us interpret stream processors as functions on streams via a concurrent "\(\run\)" function (as opposed to the sequential version from prior work \cite{Abel2012fics,Abel2016jfp}). Below, we give the type signature and code, with nested values boxed and highlighted yellow as usual.

\begin{align*}
&i,j,p:\spp_{A,B}[i,j],x:\str_A[i]\vdash y\from\run~(i, j)~(p,x)::(y:\str_B[j])\\
&y\from\run~(i, j)~(p,x)=\\
&\qquad\casee p^{\R}\,\{\get f\To h\from x^{\R}.\head\,h;\underbrace{p'\from f^{\R}.\fbox{\mathcolorbox{yellow}{\pair{h,\pair{j,p'}}}}}_{i>0\text{ assumed}};\\
&\qquad\qquad\qquad\qquad\qquad t\from\underbrace{x^{\R}.\tail t}_{i>0\text{ asserted}};\underbrace{y\from\run (i-1,j)~(p',t)}_{i,j,i>0\vdash(i-1,j)<(i,j)\text{ checked}},\\
&\qquad\qquad\qquad\pput o\To\casee y^{\W}\,\{\head\,h\To o^{\R}.\now\,h,\\
&\qquad\qquad\qquad\qquad\qquad\qquad\quad\underbrace{\tail t}_{j>0\text{ assumed}}\To p'\from\underbrace{o^{\R}.\rest p'}_{j>0\text{ asserted}};\underbrace{t\from\run~(i,j-1)~(p',x)}_{i,j,j>0\vdash(i,j-1)<(i,j)\text{ checked}}\}\}
\end{align*}

If the processor issues a ``get,'' then the head of the input stream is consumed, recursing on its tail. Otherwise, the output stream is constructed recursively, first issuing the element received from the processor. It is clear that the program terminates by lexicographic induction on $(i,j)$.
\end{exa}

\section{$\text{SAX}^\omega$}\label{sec:intermediate}

Proving termination of program reduction in the presence of infinitely deep typing derivations typically requires techniques that deviate from standard (inductive) logical relations arguments \cite{Brotherston,Derakhshan19arxiv}. As a result, we give a purely inductive process typing called $\text{SAX}^\omega$ with the judgment $\Gamma\vdash^{\omega}P::(x:A)$ (Figure \ref{tab:intermed-ty}). By dropping the arithmetic and constraint contexts, the rules $\exists{\LL}^{\omega}$ and $\forall{\R}^{\omega}$ have one premise per natural number $n$ instead of introducing a new arithmetic variable (like $\omega$-rules in arithmetic \cite{Schutte}). Moreover, the premises of ${\land}{\LL}^{\omega}$ and ${\implies}{\R}^{\omega}$ assume the closed constraint $\phi$ (which has no free arithmetic variables) holds at the meta level instead of adding it to a constraint context.

Most importantly, the call rule does not refer to a coinductively-defined auxiliary judgment, because in the absence of free arithmetic variables, the tracked size arguments decrease from some $\overline{n}$ to $\overline{n'}$ to etc. Since the lexicographic order on fixed-length natural number vectors is well-founded, this sequence necessarily terminates. To rephrase: the exact number of recursive calls is known. While this system is impractical for type checking, we can translate arithmetically closed $\text{SAX}^\infty$ derivations to $\text{SAX}^\omega$ derivations. \textcolorbf{blue}{In fact, any $\text{SAX}^\infty$ derivation can be made arithmetically closed by substituting each of its free arithmetic variables for numbers that validate (and therefore discharge) its constraints.} By trading infinitely deep derivations for infinitely wide but finitely deep ones, we may complete a logical relations argument by induction over a $\text{SAX}^\omega$ derivation. Thus, let us examine the translation theorem.

\begin{figure}[h!]
\centering
\begin{tabular}{c}
\hspace{-4.5em}$\gruleid\quad\grulecut$\\\\
\hspace{-4.5em}$\gruleoneR\quad\gruleoneL$\\\\
\hspace{-4.5em}$\gruleotimesR\quad\gruleotimesL$\\\\
\hspace{-4.5em}$\grulearrR\quad\grulearrL$\\\\
\hspace{-4.5em}$\gruleoplusR\quad\gruleoplusL$\\\\
\hspace{-4.5em}$\grulewithR\quad\grulewithL$\\\\
\hspace{-4.5em}$\gruleexistsR\quad\gruleexistsL$\\\\
\hspace{-4.5em}$\gruleforallR\quad\gruleforallL$\\\\
\hspace{-4.5em}$\textcolorbf{blue}{\gruleqR}\quad\textcolorbf{blue}{\gruleqL}$\\\\
\hspace{-4.5em}$\textcolorbf{blue}{\grulebR}\quad\textcolorbf{blue}{\grulebL}$\\\\
\hspace{-4.5em}(no rule for $\impossible$)\quad$\grulecall$
\end{tabular}
\caption{$\text{SAX}^\omega$ Typing Rules}
\label{tab:intermed-ty}
\end{figure}

\begin{thm}[Translation]
\label{thm:translation}
If $\infer*{\cdot;\cdot;\Gamma\vdash^{\overline{n}} P::(x:A)}{D}$, then $\Gamma\vdash^{\omega} P::(x:A)$.
\end{thm}
\begin{proof}

\iflongversion
By lexicographic induction on $(\overline{n},D)$, we cover the important cases. We show the proof as a transformation $\rightsquigarrow$ on derivations with $IH$ as the induction hypothesis.
\begin{enumerate}
\item When $\text{SAX}^\infty$ derivation $D$ ends in the identity or an axiomatic rule, we are done by the corresponding $\text{SAX}^\omega$ rule.
\begin{align*}
D=\infer[\id]{\cdot;\cdot;\Gamma,x:A\vdash^{\overline{n}} y^{\W}\from x^{\R}::(y:A)}{}\rightsquigarrow\gruleid
\end{align*}
\item When $\text{SAX}^\infty$ derivation $D$ ends in an invertible propositional rule with subderivation $D'$, we proceed by induction on $(\overline{n},D')$.
\begin{align*}
D=\infer[{\to}{\R}]{\cdot;\cdot;\Gamma\vdash\casee x^{\W}\,(\pair{y,z}\To P(y,z))::(x:A\to B)}{\infer*{\cdot;\cdot;\Gamma,y:A\vdash P(y,z)::(z:B)}{D'}}&\rightsquigarrow\\&\hspace{-15em}\infer[{\to}{\R}^{\omega}]{\Gamma\vdash^{\omega}\casee x^{\W}\,(\pair{y,z}\To P(y,z))::(x:A\to B)}{\infer*{\Gamma,y:A\vdash^{\omega} P(y,z)::(z:B)}{IH(\overline{n},D')}}
\end{align*}
\item When $\text{SAX}^\infty$ derivation $D$ ends in $\exists{\LL}$ or $\forall{\R}$, its subderivation $D'$ introduces a fresh arithmetic variable $i$. The $m$\textsuperscript{th} premise of the corresponding $\text{SAX}^\omega$ rules $\exists{\LL}^{\omega}$ and $\forall{\R}^{\omega}$ are fulfilled by induction on $(\overline{n},[m/i]D')$.
\begin{align*}
D=\infer[\forall{\R}]{\cdot;\cdot;\Gamma\vdash^{\overline{n}}\casee x^{\W}\,(\pair{i,y}\To P(i,y))::(x:\forall i.\,A(i))}{\infer*{i;\cdot;\Gamma\vdash^{\overline{n}} P(i,y)::(y:A(i))}{D'}}&\rightsquigarrow\\&\hspace{-20em}\infer[\forall{\R}^{\omega}]{\Gamma\vdash^{\omega}\casee x^{\W}\,(\pair{i,y}\To P(i,y))::(x:\forall i.\,A(i))}{\infer*{\Gamma\vdash^{\omega}P(m,y)::(y:A(m))~\text{for all}~m\in\mathbb{N}}{IH(\overline{n},[m/i]D')}}
\end{align*}
\item Analogously, when $\text{SAX}^\infty$ derivation $D$ ends in ${\land}{\LL}$ or ${\implies}{\R}$, its subderivation $D'$ assumes the closed constraint $\phi$. The premises of the corresponding $\text{SAX}^\omega$ rules ${\land}{\LL}^{\omega}$ and ${\implies}{\R}^{\omega}$ assume $\infer*{\cdot;\cdot\vdash\phi}{E}$, so we finish by induction on $(\overline{n},E\cdot D')$ where $\text{SAX}^\infty$ derivation $E\cdot D'$ cuts $\phi$ out of $D'$.
\begin{align*}
D=\infer[{\implies}{\R}]{\cdot;\cdot;\Gamma\vdash^{\overline{n}}\casee x^{\W}\,(\pair{\ast,y}\To P(y))::(x:\phi\implies A)}{\infer*{\cdot;\phi;\Gamma\vdash^{\overline{n}} P(y)::(y:A)}{D'}}&\rightsquigarrow\\&\hspace{-13em}\infer[{\implies}{\R}^{\omega}]{\Gamma\vdash^{\omega}\casee x^{\W}\,(\pair{\ast,y}\To P(y)::(x:\phi\implies A)}{\infer*{\Gamma\vdash^{\omega} P(y)::(y:A)~\text{if}~\infer*{\cdot;\cdot\vdash\phi}{E}}{IH(\overline{n},E\cdot D')}}
\end{align*}
\item Finally, assume $\text{SAX}^\infty$ derivation $D$ ends in the call rule with subderivation $D'$. By inversion, $D'$ ends in the $\infty$ rule with subderivation $D''$. Although $D''$ may be larger than $D$, we have some new arithmetic arguments $\overline{n'}<\overline{n}$. Thus, we are done by induction on $(\overline{n'},D'')$ then the $\text{SAX}^\omega$ call rule.
\begin{align*}
D=\infer[\text{call}]{\cdot;\cdot;\Gamma,\overline{x}:\overline{A}\vdash^{\overline{n}} y\from f~\overline{n'}~\overline{x}::(y:A)}{\cdot;\cdot\vdash\overline{n'}<\overline{n} & D'=\infer=[\infty]{\cdot;\cdot;\overline{x}:\overline{A}\vdash_{\infty}^{\overline{n'}} P_f(\overline{n'},\overline{x},y)::(y:A)}{\infer*{\cdot;\cdot;\overline{x}:\overline{A}\vdash^{\overline{n'}} P_f(\overline{n'},\overline{x},y)::(y:A)}{D''}}}&\rightsquigarrow\\&\hspace{-17em}\infer[\text{call}^{\omega}]{\Gamma,\overline{x}:\overline{A}\vdash^{\omega} y\from f~\overline{n}~\overline{x}::(y:A)}{\infer*{\overline{x}:\overline{A}\vdash^{\omega} P_f(\overline{n},\overline{x},y)::(y:A)}{IH(\overline{n'},D'')}}
\end{align*}
\end{enumerate}
\else
By lexicographic induction on $(\overline{n},D)$, we cover the important cases.
\begin{enumerate}
\item When $\text{SAX}^\infty$ derivation $D$ ends in $\exists{\LL}$ or $\forall{\R}$, its subderivation $D'$ introduces a fresh arithmetic variable $i$. The $m$\textsuperscript{th} premise of the corresponding $\text{SAX}^\omega$ rules $\exists{\LL}^{\omega}$ and $\forall{\R}^{\omega}$ are fulfilled by induction on $(\overline{n},[m/i]D')$.
\item Analogously, when $\text{SAX}^\infty$ derivation $D$ ends in ${\land}{\LL}$ or ${\implies}{\R}$, its subderivation $D'$ assumes the closed constraint $\phi$. The premises of the corresponding $\text{SAX}^\omega$ rules ${\land}{\LL}^{\omega}$ and ${\implies}{\R}^{\omega}$ assume $\infer*{\cdot;\cdot\vdash\phi}{E}$, so we finish by induction on $(\overline{n},E\cdot D')$ where $\text{SAX}^\infty$ derivation $E\cdot D'$ cuts $\phi$ out of $D'$.
\item Finally, assume $\text{SAX}^\infty$ derivation $D$ ends in the call rule with subderivation $D'$. By inversion, $D'$ ends in the $\infty$ rule with subderivation $D''$. Although $D''$ may be larger than $D$, we have some new arithmetic arguments $\overline{n'}<\overline{n}$. Thus, we are done by induction on $(\overline{n'},D'')$ then the $\text{SAX}^\omega$ call rule.
\end{enumerate}
\fi
\end{proof}

As we mentioned in the introduction, we can make the $\text{SAX}^\infty$ judgment arbitrarily rich to support more complex patterns of recursion. As long as derivations in that system can be translated to $\text{SAX}^\omega$, the logical relations argument over $\text{SAX}^\omega$ typing that we detail in Section \ref{sec:semantics} does not change. For example, consider the following additions.

\begin{enumerate}
\item\emph{Multiple blocks}: To support multiple blocks of definitions, we may simply impose the requirement that mutual recursion may not occur \emph{across} blocks. In other words, the call graph \emph{across} blocks is directed acyclic, imposing a well-founded order on definition names: $g<f$ iff $f$ calls $g$. As a result, translation of the definition $f$ may proceed by lexicographic induction on $(f,\overline{n},D)$. For example, let $f$ call $g$. If $g$ is defined in a different block than $f$, then the arithmetic arguments it applies ($\overline{n}$) may increase. Otherwise, $\overline{n}$ must decrease, since $g$ is ``equal'' to $f$ (in this order).
\item\emph{Mutual recursion with priorities}: Definitions in a block can be ordered by \emph{priority}: if $g<f$, then $f$ can call $g$ with arguments of the same size. In Example \ref{ex:odd-even}, $\odds$ calls $\evens$ with arguments of the same size but $\evens$ calls $\odds$ with arguments of lesser size. As a result, $\evens<\odds$. If $<$ is well-founded (like in this example), then translation of $f$ may proceed by lexicographic induction on $(\overline{n},f,D)$.
\end{enumerate}

\section{Operational Semantics and Termination}\label{sec:semantics}
In this section, we will give an operational semantics for \emph{configurations} of processes. Then, we will show that all $\text{SAX}^\omega$-typed processes terminate, which accounts for $\text{SAX}^\infty$ by way of Theorem \ref{thm:translation}. Program execution based on processes alone is impractical, because cut elimination only facilitates communication between two processes at a time. Thus, DeYoung et al.~\cite{DeYoung2020fscd} define programs in SAX as \emph{configurations} of simultaneously executing processes and the memory cells with which they communicate. Relatedly, the metatheory of the $\pi$-calculus must be defined up-to structural congruence to achieve a similar effect \cite{Perez2014}.

\begin{defi}[Configuration]
Let $a,b,c,\ldots$ be \emph{cell addresses} and $W:=V\,\mid\,K$. A \emph{configuration} $C$ is defined by the following grammar.

\begin{align*}
C&:=\cdot&&\text{empty configuration}\\
&\mid\,\proc{a}{P}&&\text{process}~P~\text{writing to cell addressed by}~a\\
&\mid\,\cell{a}{W}&&\text{persistent (marked with !) cell addressed by}~a~\text{with contents}~W\\
&\mid\,C,C&&\text{join of two configurations}
\end{align*}

$C$ denotes a multiset of \emph{objects} (processes and cells), so the join and empty rules form a commutative monoid. However, we also require that an address refers to at most one object in $C$. Lastly, a configuration $F$ is \emph{final} iff it only contains (persistent) cells.
\end{defi}

Now, let $\Gamma$ and $\Delta$ be contexts that associate cell addresses to types. The configuration typing judgment given in Figure \ref{tab:config-ty}, $\Gamma\vdash C::\Delta$, means that the objects in $C$ are well-typed with sources in $\Gamma$ and destinations in $\Delta$ (note that we are allowing the process typing judgment to use addresses in place of address variables). Notice that the typing rules preserve the invariant $\Gamma\subseteq\Delta$ thanks to the persistence of memory cells.
\begin{figure}[H]
\centering
\begin{center}
\begin{tabular}{c c}
\hspace{-1.5em}$\infer[\procc]{\Gamma\vdash\proc{a}{P}::(\Gamma,a:A)}{\Gamma\vdash^{\omega} P::(a:A)}$ & $\infer[{\celll}_V]{\Gamma\vdash\cell{a}{V}::(\Gamma,a:A)}{\Gamma\vdash^{\omega} a^{\W}.V::(a:A)}$\\\\
\end{tabular}
\begin{tabular}{c c c}
$\infer[{\celll}_K]{\Gamma\vdash\cell{a}{K}::(\Gamma,a:A)}{\Gamma\vdash^{\omega}\casee a^{\W}\,K::(a:A)}$ & $\infer[{\Empty}]{\Gamma\vdash\cdot::\Gamma}{}$ & $\infer[{\join}]{\Gamma\vdash C,C'::\Delta}{\Gamma\vdash C::\Gamma'&\Gamma'\vdash C'::\Delta}$
\end{tabular}
\end{center}
\caption{Configuration Typing}
\label{tab:config-ty}
\end{figure}
Configuration reduction $\to$ is given as \emph{multiset rewriting rules} \cite{Cervesato09} in Figure \ref{fig:opsem}, which replace any subset of a configuration matching the left-hand side with the right-hand side. However, $!$ indicates objects that persist across reductions. Principal cuts encountered in a configuration are resolved by passing a value to a continuation also given in Figure \ref{fig:opsem} as the relation $V\triangleright K=P$.

\begin{figure}[h]
\centering
\small
\begin{minipage}{.4\textwidth}
\begin{center}
\begin{align*}
\cell{a}{W},\proc{b}{(b^{\W}\from a^{\R})}&\to\,\cell{b}{W}\\
\proc{c}{(x\from P(x);Q(x))}&\to\\&\hspace{-8em}\proc{a}{(P(a))},\proc{c}{(Q(a))}~\text{where}~a~\text{is fresh}\\
\cell{a}{K},\proc{c}{(a^{\R}.V)}&\to\proc{c}{(V\triangleright K)}\\
\cell{a}{V},\proc{c}{(\casee a^{\R}\,K)}&\to\proc{c}{(V\triangleright K)}\\
\proc{a}{(a\from f~\overline{n}~\overline{b})}&\to\proc{a}{(P_f(\overline{n},\overline{b},a))}\\
\proc{a}{(a^{\W}.V)}&\to\,\cell{a}{V}\\
\proc{a}{(\casee a^{\W}\,K)}&\to\,\cell{a}{K}
\end{align*}
\end{center}
\end{minipage}
\qquad\vline~
\begin{minipage}{.4\textwidth}
\begin{center}
\begin{align*}
\pair{}\triangleright\pair{}\To P&=P\\
\pair{a,b}\triangleright(\pair{x,y}\To P(x,y))&=P(a,b)\\
k\,a\triangleright\{\ell\,x\To P_\ell(x)\}_{\ell\in S}&=P_k(a)\\
\pair{n,a}\triangleright(\pair{i,x}\To P(i,x))&=P(n,a)\\
\pair{\ast,a}\triangleright(\pair{\ast,x}\To P(x))&=P(a)
\end{align*}
\end{center}
\end{minipage}
\caption{Operational Semantics}
\label{fig:opsem}
\end{figure}

The first rule for $\to$ corresponds to the identity rule and copies the contents of one cell into another. \textcolorbf{red}{The second rule, which is for cut, models computing with futures \cite{Halstead85}: it allocates a new cell to be populated by the newly spawned $P$. Concurrently, $Q$ may read from said new cell, which blocks if it is not yet populated.} The third and fourth rules resolve principal cuts by passing a value to a continuation, whereas the fifth one resolves definition calls. Lastly, the final two rules perform the action of writing to a cell.

Now, we are ready to prove termination. Relatedly, refer to Das and Pfenning \cite{Das2020rast} for a proof of type safety for a session type system with arithmetic refinements. In contrast to the termination proof for base SAX \cite{DeYoung2020fscd}, we explicitly construct a model of SAX in sets of terminating configurations, also known as \emph{semantic typing} \cite{Appel2001,Kastberg2021}. This leaves open several possibilities---for example, we could reason about programs that fail to syntactically typecheck \cite{Jung2017,Dreyer} or analyze fixed points of semantic type constructors. Our approach mirrors that for natural deduction:
\begin{enumerate}
\item We define semantic types: Kripke relations \cite{Plotkin1973} over cell contents $W$ where the worlds are the (terminating) configurations to which they belong.
\item We show \emph{compatibility lemmas} that reflect the syntactic typing rules of processes, objects, and configurations in semantic typing.
\item This culminates in a fundamental theorem of the logical relation, from which termination is immediate.
\end{enumerate}
Now, let us begin with the definition of semantic type.
\begin{defi}[Semantic type]
A \emph{semantic type} $\mathscr{A},\mathscr{B},\ldots$ is a set of pairs of cell contents and final configurations, writing $F\in[W:\mathscr{A}]$ for $(W,F)\in\mathscr{A}$, such that:
\begin{itemize}
\item\label{def:weakening}\emph{Kripke monotonicity}: if $F\in[W:\mathscr{A}]$, then $F,F'\in[W:\mathscr{A}]$ for all $F'$.
\end{itemize}
Then, let $F\in[a:\mathscr{A}]\triangleq\cell{a}{W}\in F$ for some $W$ where $F\in[W:\mathscr{A}]$. We extend membership in semantic types to terminating configurations inductively in the usual way:
\[\infer{F\in\den{a:\mathscr{A}}}{F\in[a:\mathscr{A}]}\qquad\infer{C\in\den{a:\mathscr{A}}}{C'\in\den{a:\mathscr{A}}\text{ for all }C'\text{ where }C\to C'}\]
\end{defi}

Monotonicity is necessary, primarily, to prove the compatibility lemmas. In the next definition, we quickly define each semantic type in \textbf{boldface} based on its syntactic counterpart such that monotonicity is immediate.

\begin{defi}[Semantic types]
~
\begin{enumerate}
\item $F\in[\pair{}:\mathbbm{1}]$ only.
\item $F\in[\pair{a,b}:\mathscr{A}\botimes\mathscr{B}]\triangleq F\in[a:\mathscr{A}]$ and $F\in[b:\mathscr{B}]$.
\item $F\in[\pair{x,y}\To P(x,y):\mathscr{A}\blolli\mathscr{B}]\triangleq$ for all $F'\supseteq F$, if $F'\in[a:\mathscr{A}]$, then $F',\proc{b}{(P(a,b))}\in\den{b:\mathscr{B}}$ for $a$ and $b$ fresh in $F$ and $F'$, respectively.
\item $F\in[k\,a:{\boplus}\{\ell:\mathscr{A}_\ell\}_{\ell\in S}]\triangleq k\in S$ and $F\in[a:\mathscr{A}_k]$.
\item $F\in[\{\ell\,x\To P_\ell(x)\}_{\ell\in S}:{\textbf{\bwith}}\{\ell:\mathscr{A}_\ell\}_{\ell\in S}]\triangleq F,\proc{a}{(P_k(a))}\in\den{a:\mathscr{A}_k}~\text{for}~k\in S$ and for $a$ fresh in $F$.
\end{enumerate}
Assume $\sF$ is a $\mathbb{N}$-indexed semantic type and that $\phi$ is a closed constraint.
\begin{enumerate}
\item $F\in[\pair{n,a}:\bexists\mathscr{F}]\triangleq F\in[a:\mathscr{F}(n)]$.
\item $F\in[\pair{i,x}\To P(i,x):\bforall\mathscr{F}]\triangleq F,\proc{a}{(P(n,a))}\in\den{a:\mathscr{F}(n)}~\text{for all}~n\in\mathbb{N}$ and $a$ fresh in $F$.
\item $F\in[\pair{\ast,a}:\phi\bq\mathscr{A}]\triangleq\cdot;\cdot\vdash\phi$ and $F\in[a:\mathscr{A}]$.
\item $F\in[\pair{\ast,x}\To P(x):\phi\bb\mathscr{A}]\triangleq$ if $\cdot;\cdot\vdash\phi$, then $F,\proc{a}{(P(a))}\in\den{a:\mathscr{A}}$ for $a$ fresh in $F$.
\end{enumerate}
\end{defi}

Positive semantic types are defined by \emph{intension}---the contents of a particular cell---whereas negative semantic types are defined by \emph{extension}---how interacting with a continuation produces the desired result. Analogously for the $\lambda$-calculus, the semantic positive product is defined as containing pairs of terminating terms, whereas the semantic function space contains all terms that terminate under application \cite{Girard89proofstypes,Abel2016jfp}.
\iflongversion
Now, to state the compatibility lemmas, we need to define the \emph{semantic typing judgment}.
\else
Now, to state the compatibility lemmas, we need to define the \emph{semantic typing judgment}.
\fi

\begin{defi}[Semantic typing judgment]
Let $\bGamma$ and $\bDelta$ be contexts associating cell addresses to semantic types.
\begin{enumerate}
\item $F\in[\bGamma]\triangleq F\in[a:\mathscr{A}]$ for all $a:\mathscr{A}\in\bGamma$.
\item $C\in\den{\bGamma}\triangleq C\in\den{a:\mathscr{A}}$ for all $a:\mathscr{A}\in\bGamma$.
\item $\bGamma\vDash C::\bDelta\triangleq$ for all $F\in[\bGamma]$, we have $F,C\in\den{\bDelta}$.
\end{enumerate}
\end{defi}

In natural deduction, the equivalent judgment $\Gamma\vDash e:\mathscr{A}$ is defined by quantifying over all closing value substitutions $\sigma$ with domain $\Gamma$, then stating $\sigma(e)\in\mathscr{A}$. Similarly, we ask whether the configuration $C$ terminates at the desired semantic type(s) when ``closed'' by a final configuration $F$ providing all the sources from which $C$ reads. Immediately, we reproduce the standard backwards closure result.

\begin{lem}[Backward closure]
\label{thm:bwd-clo}
If for all reducts $C'$ of $C$, $\bGamma\vDash C'::\bDelta$, then $\bGamma\vDash C::\bDelta$.
\end{lem}
\iflongversion
%\begin{proof}
%Assuming $F\in[\bGamma]$, we want to show $F,C\in\den{\bDelta}$. By assumption, $F,C'\in\den{\bDelta}$. Since the only possible reduction of $F,C$ is to $F,C'$, we are done.
%\end{proof}
\fi

We are finally ready to prove a representative sample of compatibility lemmas, all of which are in Figure \ref{tab:sem-proc-ty-ii} (the dashed lines indicate that they are lemmas). Afterwards, we can tackle objects and configurations.

\begin{figure}[h!]
\centering
\begin{tabular}{c}
\hspace{-6em}$\sruleid\quad\srulecut$\\\\
\hspace{-6em}$\sruleoneR\quad\sruleoneL$\\\\
\hspace{-6em}$\sruleotimesR\quad\sruleotimesL$\\\\
\hspace{-6em}$\srulearrR\quad\srulearrL$\\\\
\hspace{-6em}$\sruleoplusR\quad\sruleoplusL$\\\\
\hspace{-6em}$\srulewithR\quad\srulewithL$\\\\
\hspace{-6em}$\sruleexistsR\quad\sruleexistsL$\\\\
\hspace{-6em}$\sruleforallR\quad\sruleforallL$\\\\
\hspace{-6em}$\textcolorbf{blue}{\sruleqR}\quad\textcolorbf{blue}{\sruleqL}$\\\\
\hspace{-6em}$\textcolorbf{blue}{\srulebR}\quad\textcolorbf{blue}{\srulebL}$\\\\
\hspace{-6em}(no rule for $\impossible$)\quad$\srulecall$
\end{tabular}
\caption{Compatibility Lemmas}
\label{tab:sem-proc-ty-ii}
\end{figure}

\begin{lem}[id]
$\bGamma,a:\mathscr{A}\vDash\proc{b}{(b^{\W}\from a^{\R})}::(b:\mathscr{A})$
\end{lem}
\begin{proof}
Assuming $F\in[\bGamma,a:\mathscr{A}]$, we want to show $F,\proc{b}{(b^{\W}\from a^{\R})}\in\den{b:\mathscr{A}}$. By assumption, there is some $\cell{a}{W}\in F$, and so because $F,\proc{b}{(b^{\W}\from a^{\R})}\to F,\cell{b}{W}$ is the only possible reduction, it suffices to show $F,\cell{b}{W}\in[b:\mathscr{A}]$ by Lemma \ref{thm:bwd-clo}, which is immediate from monotonicity.
\end{proof}

The reader may have noticed that each semantic type's definition encodes its own noninvertible rule, which makes the admissibility of rules like ${\botimes}$R immediate. Invertible rules require more effort; consider $\botimes$L and $\blolli$R below.

\begin{lem}[$\botimes$L]
\label{thm:tensL}
If $\bGamma,c:\mathscr{A}\botimes \mathscr{B},a:\mathscr{A},b:\mathscr{B}\vdash\proc{d}{(P(a,b))}::(d:\mathscr{C})$, then $\bGamma,c:\mathscr{A}\botimes\mathscr{B}\vdash\proc{d}{(\casee c^{\R}\,(\pair{x,y}\To P(x,y)))}::(d:\mathscr{C})$.
\end{lem}
\begin{proof}
Assuming $F\in[\bGamma,c:\mathscr{A}\botimes\mathscr{B}]$, we want to show that $F,\proc{d}{(\casee c^{\R}\,(\pair{x,y}\To P(x,y)))}\in\den{d:\mathscr{C}}$. Since $F\in[c:\mathscr{A}\botimes\mathscr{B}]$, we have $\cell{c}{\pair{a,b}}\in F$ where $F\in[a:\mathscr{A}]$ and $F\in[b:\mathscr{B}]$. In sum, $F\in[\bGamma,c:\mathscr{A}\botimes\mathscr{B},a:\mathscr{A},b:\mathscr{B}]$, so by the premise, $F,\proc{d}{(P(a,b))}\in\den{d:\mathscr{C}}$. Since $F,\proc{d}{(\casee c^{\R}\,(\pair{x,y}\To P(x,y)))}\to F,\proc{d}{(P(a,b))}$ is the only reduction, we are done by Lemma \ref{thm:bwd-clo}.
\end{proof}

\begin{lem}[$\blolli$R]
\label{thm:lolliR}
If $\bGamma,a:\mathscr{A}\vdash\proc{b}{(P(a,b))}::(b:\mathscr{B})$, then $\bGamma\vdash\cell{c}{(\pair{x,y}\To P(x,y))}::(c:\mathscr{A}\blolli\mathscr{B})$.
\end{lem}
\begin{proof}
Assuming $F\in[\bGamma]$, we want to show that $F,\cell{c}{(\pair{x,y}\To P(x,y))}\in[c:\mathscr{A}\blolli\mathscr{B}]$, i.e., for all $F'\supseteq F$ where $F'\in[a:\mathscr{A}]$, we have $F',\proc{b}{(P(a,b))}\in\den{b:\mathscr{B}}$. By monotonicity, $F'\in[\bGamma,a:\mathscr{A}]$, so we are done by the premise.
\end{proof}

With these compatibility lemmas in hand, we are almost ready to construct a correspondence between the syntactic typing of processes and configuration objects with the semantic typing thereof. First, we need a \emph{semantic interpretation} of (syntactic) types.

\begin{defi}[Semantic interpretation]
We define $\ned{A}_n$ by lexicographic induction on $(n,A)$ that sends arithmetically closed types to semantic ones. At a recursive type, $n$ is stepped down to allow $A$ to potentially grow larger. Note that $\lambda$ marks a meta-level anonymous function and that $\phi$ is closed.

\small\begin{align*}
\hspace{-1.5em}\ned{\mathbf{1}}_n\triangleq\mathbbm{1}\qquad\qquad\ned{A\otimes B}_n&\triangleq\ned{A}_n\botimes\ned{B}_n & \ned{A\to B}_n&\triangleq\ned{A}_n\blolli\ned{B}_n\\
\ned{{\oplus}\{\ell:A_\ell\}_{\ell\in S}}_n&\triangleq{\boplus}\{\ell:\ned{A_\ell}_n\}_{\ell\in S} & \ned{{\&}\{\ell:A_\ell\}_{\ell\in S}}_n&\triangleq{\textbf{\bwith}}\{\ell:\ned{A_\ell}_n\}_{\ell\in S}\\
\ned{\forall i.\,A(i)}_n&\triangleq\bforall(\lambda m.\,\ned{A(m)}_n) & \ned{\exists i.\,A(i)}_n&\triangleq\bexists(\lambda m.\,\ned{A(m)}_n)\\
\ned{\phi\land A}_n&\triangleq\phi\bq\ned{A}_n & \ned{\phi\implies A}_n&\triangleq\phi\bb\ned{A}_n\\
\ned{X[\overline{m}]}_0&\triangleq\emptyset & \ned{X[\overline{m}]}_{n+1}&\triangleq\ned{A_X[\overline{m}]}_n
\end{align*}
\normalsize Now, let $F\in\ned{A}\triangleq F\in\ned{A}_n$ for some $n$---intuitively, all of the (syntactic) types we have considered so far are defined by a lexicographic induction on their arithmetic indices, so $\ned{A}$ classifies configurations $C$ for which there is \emph{some} number of recursive unfoldings of $A$ that successfully establish $C$ as a member of (the denotation of) $A$ due to the aforementioned induction. $\ned{\cdot}$ is then extended to contexts $\Gamma$ and $\Delta$ in the obvious way.
\end{defi}

\begin{lem}[Semantic object typing]
\label{thm:sem-object-ty}
~
\begin{enumerate}
\item\label{thm:fund-prop-1} If $\infer*{\Gamma\vdash^{\omega} a^{\W}.V::(a:A)}{D}$, then $\ned{\Gamma}\vDash\cell{a}{V}::(a:\ned{A})$.
\item\label{thm:fund-prop-2} If $\infer*{\Gamma\vdash^{\omega}\casee a^{\W}\,K::(a:A)}{D}$, then $\ned{\Gamma}\vDash\cell{a}{K}::(a:\ned{A})$.
\item\label{thm:fund-prop-3} If $\infer*{\Gamma\vdash^{\omega} P::(a:A)}{D}$, then $\ned{\Gamma}\vDash\proc{a}{P}::(a:\ned{A})$.
\end{enumerate}
\end{lem}
\begin{proof}
\iflongversion
Part \ref{thm:fund-prop-1} follows by case analysis on $\text{SAX}^\omega$ derivation $D$ applying the relevant compatibility lemmas. For example, consider when $D$ ends in ${\otimes}{\R}^{\omega}$ (once again, representing the theorem as a translation $\rightsquigarrow$).
\begin{align*}
D=\infer[\otimes{\R}^{\omega}]{\Gamma,a:A,b:B\vdash^{\omega} c^{\W}.\pair{a,b}::(c:A\otimes B)}{}&\rightsquigarrow\\&\hspace{-11em}\infer-[\botimes{\R}]{\ned{\Gamma},a:\ned{A},b:\ned{B}\vDash\cell{c}{\pair{a,b}}::(c:\ned{A}\botimes\ned{B})}{}
\end{align*}
We prove parts \ref{thm:fund-prop-2} and \ref{thm:fund-prop-3} simultaneously by lexicographic induction on $\text{SAX}^\omega$ derivation $D$ then the part number, yielding induction hypotheses $IH_2(\text{derivation})$ and $IH_3(\text{derivation})$ indexed by part number. First, part \ref{thm:fund-prop-2} refers to part \ref{thm:fund-prop-3} on the typing subderivation for the process contained in $K$. For example, consider when $D$ ends in ${\to}{\R}^{\omega}$---upon induction, the size of the derivation decreases, allowing the part number to increase. Below, we use ellipses to indicate applications of the inductive hypothesis, even though we are not constructing syntactic typing derivations.
\begin{align*}
D=\infer[{\to}{\R}^{\omega}]{\Gamma\vdash^{\omega}\casee c^{\W}\,(\pair{y,z}\To P(y,z))::(c:A\to B)}{\infer*{\Gamma,a:A\vdash^{\omega} P(a,b)::(b:B)}{D'}}&\rightsquigarrow\\&\hspace{-15em}\infer-[{\blolli}{\R}]{\ned{\Gamma}\vDash\cell{c}{(\pair{y,z}\To P(y,z))}::(c:\ned{A}\blolli \ned{B})}{\infer*{\ned{\Gamma},a:\ned{A}\vDash\proc{b}{(P(a,b))}::(b:\ned{B})}{IH_3(D')}}
\end{align*}
In part \ref{thm:fund-prop-3}, if $P$ reads a cell with a synchronous rule, then we invoke the relevant compatibility lemma. For example, consider ${\to}{\LL}^{\omega}$.
\begin{align*}
D=\infer[{\to}{\LL}^{\omega}]{\Gamma,c:A\to B,a:A\vdash^{\omega} c^{\R}.\pair{a,b}::(b:B)}{}&\rightsquigarrow\\&\hspace{-15em}\infer-[{\blolli}{\LL}]{\ned{\Gamma},c:\ned{A}\blolli\ned{B},a:\ned{A}\vDash\proc{b}{(c^{\R}.\pair{a,b})}::(b:\ned{B})}{}
\end{align*}
However, if $P$ reads a cell with an asynchronous rule, then we invoke part \ref{thm:fund-prop-3} on the typing subderivation. For example, consider when $D$ ends in ${\otimes}{\LL}^{\omega}$---upon induction, the size of the derivation decreases, but the part number stays the same.
\begin{align*}
D=\infer[\otimes{\LL}^{\omega}]{\Gamma,c:A\otimes B\vdash^{\omega}\casee c^{\R}\,(\pair{x,y}\To P(y,z))::(d:C)}{\infer*{\Gamma,c:A\otimes B,a:A,b:B\vdash^{\omega} P(a,b)::(d:C)}{D'}}&\rightsquigarrow\\&\hspace{-26em}\infer-[\botimes{\LL}]{\ned{\Gamma},c:\ned{A}\botimes\ned{B}\vdash\proc{d}{(\casee c^{\R}\,(\pair{x,y}\To P(x,y)))}::(d:\ned{C})}{\infer*{\ned{\Gamma},c:\ned{A}\botimes\ned{B},a:\ned{A},b:\ned{B}\vdash\proc{d}{(P(a,b))}::(d:\ned{C})}{IH_3(D')}}
\end{align*}
If $P$ writes a continuation $K$, then $\proc{a}{(\casee a^{\W}\,K)}\to\cell{a}{K}$, so we invoke part \ref{thm:fund-prop-2} on $\text{SAX}^\omega$ derivation $D$ and conclude by Lemma \ref{thm:bwd-clo}. For example, we consider again the case when $D$ ends in ${\to}{\R}^{\omega}$---upon induction, the size of the derivation stays the same and the part number decreases.
\begin{align*}
D=\infer[{\to}{\R}^{\omega}]{\Gamma\vdash^{\omega}\casee c^{\W}\,(\pair{y,z}\To P(y,z))::(c:A\to B)}{\Gamma,a:A\vdash^{\omega} P(a,b)::(b:B)}&\rightsquigarrow\\&\hspace{-20em}\infer-[\text{lem.}~\ref{thm:bwd-clo}]{\ned{\Gamma}\vDash\proc{c}{(\casee c^{\W}\,(\pair{y,z}\To P(y,z)))}::(c:\ned{A}\blolli\ned{B})}{\infer*{\ned{\Gamma}\vDash\cell{c}{(\pair{y,z}\To P(y,z))}::(c:\ned{A}\blolli \ned{B})}{IH_2(D)}}
\end{align*}
Writing a value follows symmetrically since $\proc{a}{(a^{\W}.V)}\to\cell{a}{V}$, invoking part \ref{thm:fund-prop-1}. 
\begin{align*}
D=\infer[\otimes{\R}^{\omega}]{\Gamma,a:A,b:B\vdash^{\omega} c^{\W}.\pair{a,b}::(c:A\otimes B)}{}&\rightsquigarrow\\&\hspace{-20em}\infer-[\text{lem.}~\ref{thm:bwd-clo}]{\ned{\Gamma},a:\ned{A},b:\ned{B}\vDash\proc{c}{(c^{\W}.\pair{a,b})}::(c:\ned{A}\botimes\ned{B})}{\text{Part}~1~\text{on}~\infer*{\ned{\Gamma},a:\ned{A},b:\ned{B}\vDash\cell{c}{\pair{a,b}}::(c:\ned{A}\botimes\ned{B})}{D}}
\end{align*}
Lastly, identity directly invokes its semantic counterpart\ldots
\begin{align*}
D=\infer[\id^{\omega}]{\Gamma,a:A\vdash^{\omega} b^{\W}\from a^{\R}::(b:A)}{}\rightsquigarrow\infer-[{\id}]{\ned{\Gamma},a:\ned{A}\vDash\proc{b}{(b^{\W}\from a^{\R})}::(b:\ned{A})}{}
\end{align*}
\ldots whereas cut invokes the induction hypothesis (the size of the derivation decreases and the part number stays the same).
\begin{align*}
D=\infer[\cut^{\omega}]{\Gamma\vdash^{\omega} x\from P(x);Q(x)::(c:C)}{D_1\in\Gamma\vdash^{\omega} P(a)::(a:A) & D_2\in\Gamma,a:A\vdash^{\omega} Q(a)::(c:C)}&\rightsquigarrow\\&\hspace{-30em}\infer-[{\cut}]{\ned{\Gamma}\vDash\proc{c}{(x\from P;Q(x))}::(c:\ned{C})}{\infer*{\ned{\Gamma}\vDash\proc{a}{P}::(a:\ned{A})}{IH_3(D_1)} & \infer*{\ned{\Gamma},a:\ned{A}\vDash\proc{c}{(Q(a))}::(c:\ned{C})}{IH_3(D_2)}}
\end{align*}
\qedhere
\else
Part \ref{thm:fund-prop-1} follows by case analysis on $\text{SAX}^\omega$ derivation $D$ applying the relevant compatibility lemmas, like ${\botimes}{\R}$ for ${\otimes}{\R}^\omega$. We prove parts \ref{thm:fund-prop-2} and \ref{thm:fund-prop-3} simultaneously by lexicographic induction on $D$ then the part number. That is, part \ref{thm:fund-prop-2} refers to part \ref{thm:fund-prop-3} on the typing subderivation for the process contained in $K$ (like ${\to}{\R}^{\omega}$). In part \ref{thm:fund-prop-3}, if $P$ reads a cell (like ${\to}{\LL}^{\omega}$ or ${\otimes}{\LL}^{\omega}$), then we invoke the relevant compatibility lemma. If $P$ writes a continuation $K$, then $\proc{a}{(\casee a^W\,K)}\to\cell{a}{K}$, so we invoke part \ref{thm:fund-prop-2} on $D$ and conclude by Lemma \ref{thm:bwd-clo}. Writing a value follows symmetrically, invoking part \ref{thm:fund-prop-1}.
\fi
\end{proof}

Now that processes and objects have been resolved, it remains to derive the semantic configuration typing rules.

\begin{lem}[Semantic configuration typing]
\label{thm:sem-config-ty}
~
\begin{enumerate}
\item Empty: $\bGamma\vDash\cdot::\bGamma$
\item Join: If $\bGamma\vDash C::\bGamma'$ and $ \bGamma'\vDash C'::\bDelta$, then $\bGamma\vDash C,C'::\bDelta$.
\end{enumerate}
\end{lem}
\iflongversion
\begin{proof}
~
\begin{enumerate}
\item Immediate.
\item Assuming $F\in[\bGamma]$, we want to show $F,C,C'\in\den{\bDelta}$. By induction on the first premise, $F,C\in\den{\bGamma'}$, we proceed by cases. If $F,C$ is final, then we are done by the second premise. Otherwise, if $F,C\to F,C_1$ for some $C_1$, then we are done by induction hypothesis.
%so for all terminal reducts $F'$ of $F,C$, we have $F'\in[\bGamma]$. By the second premise, $F',C'\in\den{\bDelta}$. Since all reductions of $F,C,C'$ pass through $F',C'$, we are done by Lemma \ref{thm:bwd-clo}. Moreover, the compatibility lemma for cut follows from this case.
\qedhere
\end{enumerate}
\end{proof}
\fi

The previous lemmas establish the fundamental theorem, i.e., compatibility of the syntax with semantics.

\begin{thm}[Fundamental theorem]
If $\infer*{\Gamma\vdash C::\Delta}{D}$, then $\ned{\Gamma}\vDash C::\ned{\Delta}$.
\end{thm}
\begin{proof}
By induction on the configuration typing derivation $D$, the empty and join cases are discharged by Lemma \ref{thm:sem-config-ty}. The object typing cases are covered by Lemma \ref{thm:sem-object-ty}, noting that $\ned{\Gamma}$ persists across the semantic sequent due to memory cell persistence and monotonicity.
\end{proof}

The fundamental theorem of the logical relation entails termination of closed well-typed configurations, as desired.

\begin{cor}[Termination]
If $\cdot\vdash C::\Delta$, then $C$ terminates, i.e., either $C$ is final or, inductively, $C'$ terminates for all reducts $C'$ of $C$.
\end{cor}

\section{Related Work}
Our system is closely related to the sequential functional language of Lepigre and Raffalli \cite{Lepigre2019}, which utilizes circular typing derivations for a sized type system with mixed inductive-coinductive types, also avoiding continuity checking. In particular, their well-foundedness criterion on circular proofs seems to correspond to our checking that sizes decrease between recursive calls. However, they encode recursion using a fixed point combinator and use transfinite size arithmetic, both of which we avoid as we explained in the introduction. Moreover, our metatheory, which handles \emph{infinite} typing derivations (via mixed induction-coinduction at the meta level), seems to be both simpler and more general since it does not have to explicitly rule out non-circular derivations. Nevertheless, we are interested in how their innovations in polymorphism and Curry-style subtyping can be integrated into our system, especially the ability to handle programs not annotated with sizes.

\paragraph{Sized types.} Sized types are a type-oriented formulation of size-change termination \cite{Lee2001} for rewrite systems \cite{Thiemann2003,Blanqui2009}. Sized (co)inductive types \cite{Barthe2004fgpu,Blanqui,Abel2008lmcs,Abel2016jfp} gave way to sized mixed inductive-coinductive types \cite{Abel2012fics,Abel2016jfp}. In parallel, linear size arithmetic for sized inductive types \cite{Chin2001,Xi2001lics,Blaniqui2006lpar} was generalized to support coinductive types as well \cite{Sacchini14}. We present, to our knowledge, the first sized type system for a concurrent programming language as well as the first system to combine both features from above. As we mentioned in the introduction, we use unbounded quantification \cite{Vezzosi2015types} in lieu of transfinite sizes to represent (co)data of arbitrary height and depth. However, the state of the art \cite{Abel2012fics,Abel2016jfp,Chan20} supports polymorphic, higher-kinded, and dependent types, which we aim to incorporate in future work.

\paragraph{Size inference.} Our system keeps constraints implicit but arithmetic data explicit at the process level in agreement with observations made about constraint and arithmetic term reconstruction in a session-typed calculus \cite{Das20ppdp}. On the other hand, systems like $\textsf{CIC}\widehat{\phantom{}_\ell}$ \cite{Sacchini14} and $\textsf{CIC}\widehat{\ast}$ \cite{Chan20} have comprehensive \emph{size inference}, which translates recursive programs with non-sized (co)inductive types to their sized counterparts when they are well-defined. Since our view is that sized types are a mode of use of more general arithmetic refinements, we do not consider size inference at the moment.

\paragraph{Infinite and circular proofs.} Validity conditions of infinite proofs have been developed to keep cut elimination productive, which correspond to criteria like the guardedness check \cite{Baelde16csl,Berardi2017lics,Derakhshan19arxiv,Derakhshan20arxiv}. Although we use infinite typing derivations, we explicitly avoid syntactic termination checking for its non-compositionality. Nevertheless, we are interested in implementing such validity conditions as uses of sized types as future work. Relatedly, cyclic termination proofs for separation logic programs can be automated \cite{BrotherstonBornatCalcagno08,TellezBrotherston19}, although it is unclear how they could generalize to concurrent programs (in the setting of concurrent separation logic) as well as codata.

\paragraph{Session types.} Session types are inextricably linked with SAX, as it also has an asynchronous message passing interpretation \cite{Pruiksma2021jlamp}. Severi et al.~\cite{Severi16} give a mixed functional and concurrent programming language where corecursive definitions are typed with Nakano's later modality \cite{Nakano}. Since Vezzosi \cite{Vezzosi2015types} gives an embedding of the later modality and its dual into sized types, we believe that a similar arrangement can be achieved in our setting. In any case, we support recursion schemes more complex than structural (co)recursion \cite{Lindley2016icfp}.

\paragraph{$\pi$-calculi.} Certain type systems for $\pi$-calculi \cite{Kobayashi06,Padovani14,Giachino14} guarantee the eventual success of communication only if or regardless of whether processes diverge \cite{DARDHA2022100717}. \textcolorbf{blue}{Considering a configuration $C$ such that $\Gamma\vdash C::(\Gamma,a:X[n])$ where $X[i]$ is a positive coinductive type, we conjecture that $\abs{C}$, which has all constraint and arithmetic data erased, is similarly ``productive'' even if it may \emph{not} terminate. Intuitively, $C$ writes a number of cells as a function of $n$ then terminates, so $\abs{C}$ represents $C$ in the limit since $X[i]$ is positive coinductive. However, this behavior is more desirable in a message passing setting rather than in our shared memory setting.}

On the other hand, there are type systems that themselves guarantee termination---some assign numeric \emph{levels} to each channel name and restrict communication such that a measure induced by said levels decreases consistently \cite{Deng06,Demangeon10,Cristescu16}. While message passing is a different setting than ours, we are interested in the relationship between sizes and levels, if any. Other such type systems constrain the type and/or term structure; the language $\mathscr{P}$ \cite{Sangiorgi06} requires grammatical restrictions on both types and terms, the latter of which we are trying to avoid. On the other hand, the combination of linearity and a certain acyclicity condition \cite{YOSHIDA2004145} on graph types \cite{Yoshida96} is also sufficient. Our system is able to guarantee termination despite utilizing non-linear types, but it remains open how type refinements compare to graph types.

\section{Conclusion and Future Work}
We have presented a highly general concurrent language that conceives mixed inductive-coinductive programming as a mode of use of arithmetic refinements. Moreover, we prove termination via a novel logical relations argument in the presence of infinitely deep typing derivations that is mediated through infinitely wide but finitely deep (inductive) typing. There are three main points of interest for future work.
\begin{enumerate}
\item\emph{Richer types}: to mix linear \cite{notes15814}, affine linear, non-linear, etc. references to memory as well as persistent and ephemeral memory, we conjecture that moving to a type system based on adjoint logic \cite{Pruiksma2021jlamp} is appropriate. In that case, sizes could be related to the grades of the adjoint modalities \cite{Somayyajula2021tlla}. Furthermore, we are interested in generalizing to substructural, polymorphic, higher-kinded \cite{Das21arxiv}, and dependent types \cite{Cervesato96,Krishnaswami15}.
\item\emph{Implementation}: we are interested in developing a convenient surface language (perhaps a functional one \cite{Pruiksma20arxiv}) for SAX and implementing our type system, following Rast \cite{Das2020rast}, an implementation of resource-aware session types that includes arithmetic refinements. Perhaps various validity conditions of infinite proofs can be implemented as implicit uses of sized type refinements.
\item\emph{Message passing}: we would like to transport our results to the asynchronous message passing interpretation of SAX \cite{Pruiksma2021jlamp}, avoiding a technically difficult detour through asynchronous typed $\pi$-calculi \cite{DeYoung2012csl}.
\end{enumerate}

\section*{Acknowledgement}
We would like to thank Farzaneh Derakhshan, Klaas Pruiksma, Henry DeYoung, Ankush Das, and the anonymous reviewers for helpful discussion and suggestions regarding the contents of this paper.

\bibliographystyle{alpha}
\bibliography{refs}

\end{document}

%% file: defs.tex
\DeclarePairedDelimiter{\pair}{\langle}{\rangle}
\DeclarePairedDelimiter{\den}{\llbracket}{\rrbracket}
\DeclarePairedDelimiter{\abs}{\lvert}{\rvert}

\newcommand{\To}{\Rightarrow}
\newcommand{\V}{\mathcal{V}}
\newcommand{\CC}{\mathcal{C}}

\DeclareMathOperator{\procc}{proc}
\newcommand{\proc}[2]{\procc #1\,#2}

\DeclareMathOperator{\celll}{!cell}
\newcommand{\cell}[2]{\celll #1\,#2}

\newcommand{\from}{\leftarrow}

\DeclareMathOperator{\append}{append}
\DeclareMathOperator{\partition}{partition}
\DeclareMathOperator{\qsort}{qsort}
\DeclareMathOperator{\nil}{nil}
\DeclareMathOperator{\cons}{cons}

\newcommand{\listT}{\mathrm{list}}

\newcommand{\botimes}{\pmb{\bm{\otimes}}}
\newcommand{\boplus}{\pmb{\bm{\oplus}}}
\newcommand{\bexists}{\pmb{\bm{\exists}}}
\newcommand{\bforall}{\pmb{\bm{\forall}}}
\newcommand{\bq}{\pmb{\bm{\land}}}
\newcommand{\bb}{\pmb{\bm{\implies}}}

\newcommand{\blolli}{\pmb{\bm{\to}}}

\newcommand{\bGamma}{\bm{\Gamma}}
\newcommand{\bDelta}{\bm{\Delta}}

\newcommand{\sF}{\mathscr{F}}

\DeclarePairedDelimiter{\ned}{\llparenthesis}{\rrparenthesis}

\DeclareMathOperator{\impossible}
{\mathbf{impossible}}
\DeclareMathOperator{\casee}{\mathbf{case}}

\DeclareMathOperator{\str}{stream}
\DeclareMathOperator{\spp}{sp}
\DeclareMathOperator{\head}{head}
\DeclareMathOperator{\tail}{tail}
\DeclareMathOperator{\get}{get}
\DeclareMathOperator{\pput}{put}
\DeclareMathOperator{\now}{now}
\DeclareMathOperator{\rest}{rest}
\DeclareMathOperator{\zero}{zero}
\DeclareMathOperator{\succc}{succ}
\DeclareMathOperator{\nat}{nat}
\DeclareMathOperator{\eat}{eat}

\DeclareMathOperator{\evens}{evens}
\DeclareMathOperator{\odds}{odds}

\DeclareMathOperator{\run}{run}

\newcommand{\one}{\mathbf{1}}
\newcommand{\bwith}{\bm{\&}}

\DeclareMathOperator{\id}{id}
\DeclareMathOperator{\cut}{cut}

\DeclareMathOperator{\R}{R}
\DeclareMathOperator{\LL}{L}
\DeclareMathOperator{\W}{W}

\DeclareMathOperator{\Empty}{empty}
\DeclareMathOperator{\join}{join}

\DeclareMathOperator{\lfair}{lfair}

\DeclareMathOperator{\san}{proj}

\newcommand{\ruleid}{
\infer[\id]{\V;\CC;\Gamma,x:A\vdash^{\overline{e}} y^{\W}\from x^{\R}::(y:A)}{}
}

\newcommand{\gruleid}{
\infer[\id^{\omega}]{\Gamma,x:A\vdash^{\omega} y^{\W}\from x^{\R}::(y:A)}{}
}

\newcommand{\sruleid}{
\infer-[{\id}]{\bGamma,a:\mathscr{A}\vDash\proc{b}{(b^{\W}\from a^{\R})}::(b:\mathscr{A})}{}
}

\newcommand{\rulecut}{
\infer[\cut]{\V;\CC;\Gamma\vdash^{\overline{e}} x\from P(x);Q(x)::(z:C)}{\V;\CC;\Gamma\vdash^{\overline{e}} P(x)::(x:A) & \V;\CC;\Gamma,x:A\vdash^{\overline{e}} Q(x)::(z:C)}
}

\newcommand{\grulecut}{
\infer[\cut^{\omega}]{\Gamma\vdash^{\omega} x\from P(x);Q(x)::(z:C)}{\Gamma\vdash^{\omega} P(x)::(x:A) & \Gamma,x:A\vdash^{\omega} Q(x)::(z:C)}
}

\newcommand{\srulecut}{
\infer-[{\cut}]{\bGamma\vDash\proc{c}{(x\from P;Q(x))}::(c:\mathscr{C})}{\bGamma\vDash\proc{a}{P}::(a:\mathscr{A}) & \bGamma,a:A\vDash\proc{c}{(Q(a))}::(c:\mathscr{C})}
}

\newcommand{\ruleoneR}{
\infer[\one{\R}]{\V;\CC;\Gamma\vdash^{\overline{e}} x^{\W}.\pair{}::(x:\one)}{}
}

\newcommand{\gruleoneR}{
\infer[\one{\R}^{\omega}]{\Gamma\vdash^{\omega} x^{\W}.\pair{}::(x:\one)}{}
}

\newcommand{\sruleoneR}{
\infer-[\mathbbm{1}{\R}]{\bGamma\vDash\cell{a}{\pair{}}::(a:\mathbbm{1})}{}
}

\newcommand{\ruleoneL}{
\infer[\one{\LL}]{\V;\CC;\Gamma,x:\one\vdash^{\overline{e}}\casee x^{\R}\,(\pair{}\To P)::(z:C)}{\V;\CC;\Gamma,x:\one\vdash^{\overline{e}} P::(z:C)}
}

\newcommand{\gruleoneL}{
\infer[\one{\LL}^{\omega}]{\Gamma,x:\one\vdash^{\omega}\casee x^{\R}\,(\pair{}\To P)::(z:C)}{\Gamma,x:\one\vdash^{\omega} P::(z:C)}
}

\newcommand{\sruleoneL}{
\infer-[\mathbbm{1}{\LL}]{\bGamma,a:\mathbbm{1}\vDash\proc{b}{(\casee a^{\R}\,(\pair{}\To P))}::(b:\mathscr{C})}{\bGamma,a:\mathbbm{1}\vDash\proc{b}{P}::(b:\mathscr{C})}
}

\newcommand{\ruleotimesR}{
\infer[\otimes{\R}]{\V;\CC;\Gamma,y:A,z:B\vdash^{\overline{e}} x^{\W}.\pair{y,z}::(x:A\otimes B)}{}
}

\newcommand{\gruleotimesR}{
\infer[\otimes{\R}^{\omega}]{\Gamma,y:A,z:B\vdash^{\omega} x^{\W}.\pair{y,z}::(x:A\otimes B)}{}
}

\newcommand{\sruleotimesR}{
\infer-[\botimes{\R}]{\bGamma,a:\mathscr{A},b:\mathscr{B}\vDash\cell{c}{\pair{a,b}}::(c:\mathscr{A}\botimes\mathscr{B})}{}
}

\newcommand{\ruleotimesL}{
\infer[\otimes{\LL}]{\V;\CC;\Gamma,x:A\otimes B\vdash^{\overline{e}}\casee x^{\R}\,(\pair{y,z}\To P(y,z))::(w:C)}{\V;\CC;\Gamma,x:A\otimes B,y:A,z:B\vdash^{\overline{e}} P(y,z)::(w:C)}
}

\newcommand{\gruleotimesL}{
\infer[\otimes{\LL}^{\omega}]{\Gamma,x:A\otimes B\vdash^{\omega}\casee x^{\R}\,(\pair{y,z}\To P(y,z))::(w:C)}{\Gamma,x:A\otimes B,y:A,z:B\vdash^{\omega} P(y,z)::(w:C)}
}

\newcommand{\sruleotimesL}{
\infer-[\botimes{\LL}]{\bGamma,c:\mathscr{A}\botimes\mathscr{B}\vdash\proc{d}{(\casee c^{\R}\,(\pair{x,y}\To P(x,y)))}::(d:\mathscr{D})}{\bGamma,c:\mathscr{A}\botimes \mathscr{B},a:\mathscr{A},b:\mathscr{B}\vdash\proc{d}{(P(a,b))}::(d:\mathscr{D})}
}

\newcommand{\rulearrR}{
\infer[{\to}{\R}]{\V;\CC;\Gamma\vdash^{\overline{e}}\casee x^{\W}\,(\pair{y,z}\To P(y,z))::(x:A\to B)}{\V;\CC;\Gamma,y:A\vdash^{\overline{e}} P(y,z)::(z:B)}
}

\newcommand{\grulearrR}{
\infer[{\to}{\R}^{\omega}]{\Gamma\vdash^{\omega}\casee x^{\W}\,(\pair{y,z}\To P(y,z))::(x:A\to B)}{\Gamma,y:A\vdash^{\omega} P(y,z)::(z:B)}
}

\newcommand{\srulearrR}{
\infer-[{\blolli}{\R}]{\bGamma\vDash\cell{c}{(\pair{x,y}\To P(x,y))}::(c:\mathscr{A}\blolli \mathscr{B})}{\bGamma,a:\mathscr{A}\vDash\proc{b}{(P(a,b))}::(b:\mathscr{B})}
}

\newcommand{\rulearrL}{
\infer[{\to}{\LL}]{\V;\CC;\Gamma,x:A\to B,y:A\vdash^{\overline{e}} x^{\R}.\pair{y,z}::(z:B)}{}
}

\newcommand{\grulearrL}{
\infer[{\to}{\LL}^{\omega}]{\Gamma,x:A\to B,y:A\vdash^{\omega} x^{\R}.\pair{y,z}::(z:B)}{}
}

\newcommand{\srulearrL}{
\infer-[{\blolli}{\LL}]{\bGamma,c:\mathscr{A}\blolli\mathscr{B},a:\mathscr{A}\vDash\proc{b}{(c^{\R}.\pair{a,b})}::(b:\mathscr{B})}{}
}

\newcommand{\ruleoplusR}{
\infer[\oplus{\R}]{\V;\CC;\Gamma,y:A_k\vdash^{\overline{e}} x^{\W}.k\,y::(x:\oplus\{\ell:A_\ell\}_{\ell\in S})}{k\in S}
}

\newcommand{\gruleoplusR}{
\infer[\oplus{\R}^{\omega}]{\Gamma,y:A_k\vdash^{\omega} x^{\W}.k\,y::(x:\oplus\{\ell:A_\ell\}_{\ell\in S})}{k\in S}
}

\newcommand{\sruleoplusR}{
\infer-[{\boplus}{\R}]{\bGamma,a:\mathscr{A}_k\vDash\cell{b}{(k\,a)}::(b:\oplus\{\ell:\mathscr{A_\ell}\}_{\ell\in S})}{}
}

\newcommand{\ruleoplusL}{
\infer[\oplus{\LL}]{\V;\CC;\Gamma,x:\oplus\{\ell:A_\ell\}_{\ell\in S}\vdash^{\overline{e}}\casee x^{\R}\,\{\ell\,y\To P_\ell(y)\}_{\ell\in S}::(z:C)}{\{\V;\CC;\Gamma,x:\oplus\{\ell:A_\ell\}_{\ell\in S},y:A_\ell\vdash^{\overline{e}} P(y)::(z:C)\}_{\ell\in S}}
}

\newcommand{\gruleoplusL}{
\infer[\oplus{\LL}^{\omega}]{\Gamma,x:\oplus\{\ell:A_\ell\}_{\ell\in S}\vdash^{\omega}\casee x^{\R}\,\{\ell\,y\To P_\ell(y)\}_{\ell\in S}::(z:C)}{\{\Gamma,x:\oplus\{\ell:A_\ell\}_{\ell\in S},y:A_\ell\vdash^{\omega} P(y)::(z:C)\}_{\ell\in S}}
}

\newcommand{\sruleoplusL}{
\infer-[\boplus{\LL}]{\bGamma,b:\boplus\{\ell:\mathscr{A}_\ell\}_{\ell\in S}\vDash\proc{c}{(\casee b^{\R}\,\{\ell\,x\To P_\ell(x)\}_{\ell\in S})}::(c:\mathscr{C})}{\{\bGamma,b:\boplus\{\ell:\mathscr{A}_\ell\}_{\ell\in S},a:\mathscr{A}_k\vDash\proc{c}{(P_k(a))}::(c:\mathscr{C})\}_{k\in S}}
}

\newcommand{\rulewithR}{
\infer[\&{\R}]{\V;\CC;\Gamma\vdash^{\overline{e}}\casee x^{\W}\,\{\ell\,y\To P_\ell(y)\}_{\ell\in S}::(x:\&\{\ell:A_\ell\}_{\ell\in S})}{\{\V;\CC;\Gamma\vdash^{\overline{e}} P(y)::(y:A_\ell)\}_{\ell\in S}}
}

\newcommand{\grulewithR}{
\infer[\&{\R}^{\omega}]{\Gamma\vdash^{\omega}\casee x^{\W}\,\{\ell\,y\To P_\ell(y)\}_{\ell\in S}::(x:\&\{\ell:A_\ell\}_{\ell\in S})}{\{\Gamma\vdash^{\omega} P(y)::(y:A_\ell)\}_{\ell\in S}}
}

\newcommand{\srulewithR}{
\infer-[\textbf{\bwith}{\R}]{\bGamma\vDash\cell{b}{\{\ell\,x\To P_\ell(x)\}_{\ell\in S}}::(b:\textbf{\bwith}\{\ell:\mathscr{A}_\ell\}_{\ell\in S})}{\{\bGamma\vDash\proc{a}{(P_\ell(a))}::(a:\mathscr{A}_\ell)\}_{\ell\in S}}
}

\newcommand{\rulewithL}{
\infer[\&{\LL}]{\V;\CC;\Gamma,x:\&\{\ell:A_\ell\}_{\ell\in S}\vdash^{\overline{e}} x^{\R}.k\,y::(y:A_k)}{k\in S}
}

\newcommand{\grulewithL}{
\infer[\&{\LL}^{\omega}]{\Gamma,x:\&\{\ell:A_\ell\}_{\ell\in S}\vdash^{\omega} x^{\R}.k\,y::(y:A_k)}{k\in S}
}

\newcommand{\srulewithL}{
\infer-[\textbf{\bwith}{\LL}]{\bGamma,b:\textbf{\&}\{\ell:\mathscr{A_\ell}\}_{\ell\in S}\vDash\proc{b}{(c^{\R}.k\,a)}::(a:\mathscr{A_k})}{}
}

\newcommand{\ruleexistsR}{
\infer[\exists{\R}]{\V;\CC;\Gamma,y:A(e)\vdash^{\overline{e}} x^{\W}.\pair{e,y}::(x:\exists i.\,A(i))}{\V;\CC\vdash e}
}

\newcommand{\gruleexistsR}{
\infer[\exists{\R}^{\omega}]{\Gamma,y:A(n)\vdash^{\omega} x^{\W}.\pair{n,y}::(x:\exists i.\,A(n))}{}
}

\newcommand{\sruleexistsR}{
\infer-[\bexists{\R}]{\bGamma,a:\mathscr{F}(n)\vDash\cell{b}{\pair{n,a}}::(b:\bexists\mathscr{F})}{}
}

\newcommand{\ruleexistsL}{
\infer[\exists{\LL}]{\V;\CC;\Gamma,x:\exists i.\,A(i)\vdash^{\overline{e}}\casee x^{\R}\,(\pair{i,y}\To P(i,y))::(z:C)}{\V,i;\CC;\Gamma,x:\exists i.\,A(i),y:A(i)\vdash^{\overline{e}} P(i,y)::(z:C)}
}

\newcommand{\gruleexistsL}{
\infer[\exists{\LL}^{\omega}]{\Gamma,x:\exists i.\,A(i)\vdash^{\omega}\casee x^{\R}\,(\pair{i,y}\To P(i,y))::(z:C)}{\Gamma,x:\exists i.\,A(i),y:A(n)\vdash^{\omega} P(n,y)::(z:C)~\text{for all}~n\in\mathbb{N}}
}

\newcommand{\sruleexistsL}{
\infer-[\bexists{\LL}]{\bGamma,b:\bexists\mathscr{F}\vDash\proc{c}{(\casee b^{\R}\,(\pair{i,x}\To P(i,x)))}::(c:\mathscr{C})}{\{\bGamma,b:\bexists\mathscr{F},a:\mathscr{F}(n)\vDash\proc{c}{(P(n,a))}::(c:\mathscr{C})\}_{n\in\mathbb{N}}}
}

\newcommand{\ruleforallR}{
\infer[\forall{\R}]{\V;\CC;\Gamma\vdash^{\overline{e}}\casee x^{\W}\,(\pair{i,y}\To P(i,y))::(x:\forall i.\,A(i))}{\V,i;\CC;\Gamma\vdash^{\overline{e}} P(i,y)::(y:A(i))}
}

\newcommand{\gruleforallR}{
\infer[\forall{\R}^{\omega}]{\Gamma\vdash^{\omega}\casee x^{\W}\,(\pair{i,y}\To P(i,y))::(x:\forall i.\,A(i))}{\Gamma\vdash^{\omega} P(n,y)::(y:A(n))~\text{for all}~n\in\mathbb{N}}
}

\newcommand{\ruleforallL}{
\infer[\forall{\LL}]{\V;\CC;\Gamma,x:\forall i.\,A(i)\vdash^{\overline{e}} x^{\R}.\pair{e,y}::(y:A(e))}{\V;\CC\vdash e}
}

\newcommand{\gruleforallL}{
\infer[\forall{\LL}^{\omega}]{\Gamma,x:\forall i.\,A(i)\vdash^{\omega} x^{\R}.\pair{n,y}::(y:A(n))}{}
}

\newcommand{\sruleforallR}{
\infer-[\bforall{\R}]{\bGamma\vDash\cell{b}{(\pair{i,x}\To P(i,x))}::(b:\bforall\mathscr{F})}{\{\bGamma\vDash\proc{a}{(P(n,a))}::(a:\mathscr{F}(n))\}_{n\in\mathbb{N}}}
}

\newcommand{\sruleforallL}{
\infer-[\bforall{\LL}]{\bGamma,b:\bforall\mathscr{F}\vDash\proc{a}{(b^{\R}.\pair{n,a})}::(a:\mathscr{F}(n))}{}
}

\newcommand{\ruleqR}{
\infer[\land{\R}]{\V;\CC;\Gamma,y:A\vdash^{\overline{e}}x^{\W}.\pair{\ast,y}::(x:\phi\land A)}{\V;\CC\vdash\phi}
}

\newcommand{\gruleqR}{
\infer[\land{\R}^{\omega}]{\Gamma,y:A\vdash^{\omega} x^{\W}.\pair{\ast,y}::(x:\phi\land A)}{\cdot;\cdot\vdash\phi}
}

\newcommand{\sruleqR}{
\infer-[{\bq}{\R}]{\bGamma,b:A\vDash\cell{a}{\pair{\ast,b}}::(a:\phi\bq\mathscr{A})}{\cdot;\cdot\vdash\phi}
}

\newcommand{\ruleqL}{
\infer[{\land}{\LL}]{\V;\CC;\Gamma,x:\phi\land A\vdash^{\overline{e}}\casee x^{\R}\,(\pair{\ast,y}\To P(y))::(z:C)}{\V;\CC,\phi;\Gamma,x:\phi\land A,y:A\vdash^{\overline{e}} P(y)::(z:C)}
}

\newcommand{\gruleqL}{
\infer[{\land}{\LL}^{\omega}]{\Gamma,x:\phi\land A\vdash^{\omega}\casee x^{\R}\,(\pair{\ast,y}\To P(y))::(z:C)}{\Gamma,x:\phi\land A,y:A\vdash^{\omega} P(y)::(z:C)~\text{if}~\cdot;\cdot\vdash\phi}
}

\newcommand{\sruleqL}{
\infer-[{\bq}{\LL}]{\bGamma,a:\phi\bq\mathscr{A}\vDash\proc{a}{(\casee a^{\R}\,(\pair{\ast,y}\To P(y)))}::(c:\mathscr{C})}{\bGamma,a:\phi\bq\mathscr{A},b:\mathscr{A}\vdash\proc{a}{(P(b))}::(c:\mathscr{C})~~\text{if}~\cdot;\cdot\vdash\phi}
}

\newcommand{\rulebR}{
\infer[{\implies}{\R}]{\V;\CC;\Gamma\vdash^{\overline{e}}\casee x^{\W}\,(\pair{\ast,y}\To P(y))::(x:\phi\implies A)}{\V;\CC,\phi;\Gamma\vdash^{\overline{e}} P(y)::(y:A)}
}

\newcommand{\grulebR}{
\infer[{\implies}{\R}^{\omega}]{\Gamma\vdash^{\omega}\casee x^{\W}\,(\pair{\ast,y}\To P(y))::(x:\phi\implies A)}{\Gamma\vdash^{\omega} P(y)::(y:A)~\text{if}~\cdot;\cdot\vdash\phi}
}

\newcommand{\srulebR}{
\infer-[{\bb}{\R}]{\bGamma\vDash\cell{a}{(\pair{\ast,y}\To P(y))}::(a:\phi\bb\mathscr{A})}{\bGamma\vDash\proc{b}{(P(b))}::(b:\mathscr{A})~\text{if}~\cdot;\cdot\vdash\phi}
}

\newcommand{\rulebL}{
\infer[{\implies}{\LL}]{\V;\CC;\Gamma,x:\phi\implies A\vdash^{\overline{e}}x^{\R}.\pair{\ast,y}::(y:A)}{\V;\CC\vdash\phi}
}

\newcommand{\grulebL}{
\infer[{\implies}{\LL}^{\omega}]{\Gamma,x:\phi\implies A\vdash^{\omega}x^{\R}.\pair{\ast,y}::(y:A)}{\cdot;\cdot\vdash\phi}
}

\newcommand{\srulebL}{
\infer-[{\bb}{\LL}]{\bGamma,a:\phi\bb\mathscr{A}\vDash\proc{b}{(a^{\R}.\pair{\ast,b})}::(b:\mathscr{A})}{\cdot;\cdot\vdash\phi}
}

\newcommand{\rulecall}{
\infer[\text{call}]{\V;\CC;\Gamma,\overline{x}:\overline{A}\vdash^{\overline{e}} y\from f~\overline{e'}~\overline{x}::(y:A)}{\V;\CC\vdash\overline{e'}<\overline{e} & y\from f~\overline{i}~\overline{x}=P_f(\overline{i},\overline{x},y) & \V;\CC;\overline{x}:\overline{A}\vdash_\infty^{\overline{e'}} P_f(\overline{e'},\overline{x},y)::(y:A)}
}

\newcommand{\rulein}{
\infer=[\infty]{\V;\CC;\Gamma\vdash_\infty^{\overline{e}}P::(y:A)}{\V;\CC;\Gamma\vdash^{\overline{e}}P::(y:A)}
}

\newcommand{\grulecall}{
\infer[\text{call}^{\omega}]{\Gamma,\overline{x}:\overline{A}\vdash^{\omega} y\from f~\overline{n}~\overline{x}::(y:A)}{y\from f~\overline{i}~\overline{x}=P_f(\overline{i},\overline{x},y) & \overline{x}:\overline{A}\vdash^{\omega} P_f(\overline{n},\overline{x},y)::(y:A)}
}

\newcommand{\srulecall}{
\infer-[\text{call}]{\bGamma,\overline{b}:\overline{\mathscr{A}}\vDash\proc{a}{(a\from f~\overline{n}~\overline{b})}::(a:\mathscr{A})}{y\from f~\overline{i}~\overline{x}=P_f(\overline{i},\overline{x},y) & \overline{b}:\overline{\mathscr{A}}\vDash\proc{a}{(P_f(\overline{n},\overline{b},a))}::(a:\mathscr{A})}
}

\renewcommand{\implies}{\Rightarrow}

\DeclareMathOperator{\later}{later}

\newcommand{\mathcolorbox}[2]{\colorbox{#1}{$\displaystyle #2$}}

%% file: main.bbl
\newcommand{\etalchar}[1]{$^{#1}$}
\begin{thebibliography}{SPTDC16}

\bibitem[Abe]{StreamWorkshop}
Andreas Abel.
\newblock Productive infinite objects via copatterns and sized types in agda.

\bibitem[Abe08]{Abel2008lmcs}
Andreas Abel.
\newblock {Semi-continuous Sized Types and Termination}.
\newblock {\em {Logical Methods in Computer Science}}, {Volume 4, Issue 2},
  April 2008.

\bibitem[Abe12]{Abel2012fics}
Andreas Abel.
\newblock Type-based termination, inflationary fixed-points, and mixed
  inductive-coinductive types.
\newblock In Dale Miller and Zolt{\'{a}}n {\'{E}}sik, editors, {\em Proceedings
  8th Workshop on Fixed Points in Computer Science, {FICS} 2012, Tallinn,
  Estonia, 24th March 2012}, volume~77 of {\em {EPTCS}}, pages 1--11, 2012.

\bibitem[AP16]{Abel2016jfp}
Andreas Abel and Brigitte Pientka.
\newblock {Well-founded recursion with copatterns and sized types}.
\newblock {\em Journal of Functional Programming}, 26:e2, 2016.

\bibitem[App01]{Appel2001}
Andrew~W. Appel.
\newblock Foundational proof-carrying code.
\newblock In {\em Proceedings of the 16th Annual IEEE Symposium on Logic in
  Computer Science}, LICS '01, page 247, USA, 2001. IEEE Computer Society.

\bibitem[Bas18]{Basold2018}
Henning Basold.
\newblock {\em Mixed inductive-coinductive reasoning: types, programs and
  logic}.
\newblock PhD thesis, Radboud University, 2018.

\bibitem[BBC08]{BrotherstonBornatCalcagno08}
James Brotherston, Richard Bornat, and Cristiano Calcagno.
\newblock Cyclic proofs of program termination in separation logic.
\newblock In {\em Proceedings of {POPL}-35}, pages 101--112. ACM, 2008.

\bibitem[BDS16]{Baelde16csl}
David Baelde, Amina Doumane, and Alexis Saurin.
\newblock {Infinitary Proof Theory: the Multiplicative Additive Case}.
\newblock In Jean-Marc Talbot and Laurent Regnier, editors, {\em 25th EACSL
  Annual Conference on Computer Science Logic (CSL 2016)}, volume~62 of {\em
  Leibniz International Proceedings in Informatics (LIPIcs)}, pages
  42:1--42:17, Dagstuhl, Germany, 2016. Schloss Dagstuhl--Leibniz-Zentrum fuer
  Informatik.

\bibitem[BFG{\etalchar{+}}04]{Barthe2004fgpu}
Gilles Barthe, Maria~Jo{\~{a}}o Frade, Eduardo Gim{\'{e}}nez, Lu{\'{\i}}s
  Pinto, and Tarmo Uustalu.
\newblock {Type-based termination of recursive definitions}.
\newblock {\em Math. Struct. Comput. Sci.}, 14(1):97--141, 2004.

\bibitem[Bla04]{Blanqui}
Fr{\'e}d{\'e}ric Blanqui.
\newblock A type-based termination criterion for dependently-typed higher-order
  rewrite systems.
\newblock In Vincent van Oostrom, editor, {\em Rewriting Techniques and
  Applications}, pages 24--39, Berlin, Heidelberg, 2004. Springer Berlin
  Heidelberg.

\bibitem[BR06]{Blaniqui2006lpar}
Fr{\'e}d{\'e}ric Blanqui and Colin Riba.
\newblock Combining typing and size constraints for checking the termination of
  higher-order conditional rewrite systems.
\newblock In Miki Hermann and Andrei Voronkov, editors, {\em Logic for
  Programming, Artificial Intelligence, and Reasoning}, pages 105--119, Berlin,
  Heidelberg, 2006. Springer Berlin Heidelberg.

\bibitem[BR09]{Blanqui2009}
Fr{\'e}d{\'e}ric Blanqui and Cody Roux.
\newblock {On the relation between sized-types based termination and semantic
  labelling}.
\newblock In {\em {18th EACSL Annual Conference on Computer Science Logic - CSL
  09}}, Coimbra, Portugal, September 2009.
\newblock Full version.

\bibitem[Bro05]{Brotherston}
James Brotherston.
\newblock {Cyclic Proofs for First-Order Logic with Inductive Definitions}.
\newblock In Bernhard Beckert, editor, {\em Automated Reasoning with Analytic
  Tableaux and Related Methods}, pages 78--92, Berlin, Heidelberg, 2005.
  Springer Berlin Heidelberg.

\bibitem[BT17]{Berardi2017lics}
Stefano Berardi and Makoto Tatsuta.
\newblock {Equivalence of Inductive Definitions and Cyclic Proofs under
  Arithmetic}.
\newblock In {\em 2017 32nd Annual ACM/IEEE Symposium on Logic in Computer
  Science (LICS)}, pages 1--12, 2017.

\bibitem[CH16]{Cristescu16}
Ioana Cristescu and Daniel Hirschkoff.
\newblock Termination in a $\pi$-calculus with subtyping.
\newblock {\em Mathematical Structures in Computer Science}, 26(8):1395–1432,
  2016.

\bibitem[CK01]{Chin2001}
Wei-Ngan Chin and Siau-Cheng Khoo.
\newblock Calculating sized types.
\newblock {\em Higher-Order and Symbolic Computation}, 14(2):261--300, 2001.

\bibitem[CLB23]{Chan20}
Jonathan Chan, Yufeng Li, and William~J. Bowman.
\newblock {Is sized typing for Coq practical?}
\newblock {\em Journal of Functional Programming}, 33:e1, 2023.

\bibitem[CP96]{Cervesato96}
Iliano Cervesato and Frank Pfenning.
\newblock A linear logical framework.
\newblock In {\em Proceedings of the 11th Annual IEEE Symposium on Logic in
  Computer Science}, LICS '96, page 264, USA, 1996. IEEE Computer Society.

\bibitem[CP10]{Caires10concur}
Lu{\'\i}s Caires and Frank Pfenning.
\newblock {Session Types as Intuitionistic Linear Propositions}.
\newblock In {\em Proceedings of the 21st International Conference on
  Concurrency Theory (CONCUR 2010)}, pages 222--236, Paris, France, August
  2010. Springer LNCS 6269.

\bibitem[CS09]{Cervesato09}
Iliano Cervesato and Andre Scedrov.
\newblock {Relating state-based and process-based concurrency through linear
  logic}.
\newblock {\em Information and Computation}, 207(10):1044--1077, 2009.
\newblock Special issue: 13th Workshop on Logic, Language, Information and
  Computation (WoLLIC 2006).

\bibitem[DA09]{Danielsson09}
Nils~Anders Danielsson and Thorsten Altenkirch.
\newblock Mixing induction and coinduction, 2009.

\bibitem[Dag21]{Dagnino}
Francesco Dagnino.
\newblock {Foundations of regular coinduction}.
\newblock {\em {Logical Methods in Computer Science}}, {Volume 17, Issue 4},
  October 2021.

\bibitem[DCPT12]{DeYoung2012csl}
Henry DeYoung, Lu{\'i}s Caires, Frank Pfenning, and Bernardo Toninho.
\newblock {Cut Reduction in Linear Logic as Asynchronous Session-Typed
  Communication}.
\newblock In Patrick C{\'e}gielski and Arnaud Durand, editors, {\em Computer
  Science Logic (CSL'12) - 26th International Workshop/21st Annual Conference
  of the EACSL}, volume~16 of {\em Leibniz International Proceedings in
  Informatics (LIPIcs)}, pages 228--242, Dagstuhl, Germany, 2012. Schloss
  Dagstuhl--Leibniz-Zentrum fuer Informatik.

\bibitem[DDMP21]{Das21arxiv}
Ankush Das, Henry DeYoung, Andreia Mordido, and Frank Pfenning.
\newblock {Subtyping on Nested Polymorphic Session Types}, 2021.

\bibitem[DHS10]{Demangeon10}
Romain Demangeon, Daniel Hirschkoff, and Davide Sangiorgi.
\newblock Termination in impure concurrent languages.
\newblock In Paul Gastin and Fran{\c{c}}ois Laroussinie, editors, {\em CONCUR
  2010 - Concurrency Theory}, pages 328--342, Berlin, Heidelberg, 2010.
  Springer Berlin Heidelberg.

\bibitem[DP19]{Derakhshan19arxiv}
Farzaneh Derakhshan and Frank Pfenning.
\newblock {Circular Proofs as Session-Typed Processes: {A} Local Validity
  Condition}.
\newblock {\em CoRR}, abs/1908.01909, August 2019.

\bibitem[DP20a]{Das2020rast}
Ankush Das and Frank Pfenning.
\newblock {Rast: A Language for Resource-Aware Session Types}, 2020.

\bibitem[DP20b]{Das20concur}
Ankush Das and Frank Pfenning.
\newblock {Session Types with Arithmetic Refinements}.
\newblock In Igor Konnov and Laura Kov{\'a}cs, editors, {\em 31st International
  Conference on Concurrency Theory (CONCUR 2020)}, volume 171 of {\em Leibniz
  International Proceedings in Informatics (LIPIcs)}, pages 13:1--13:18,
  Dagstuhl, Germany, 2020. Schloss Dagstuhl--Leibniz-Zentrum f{\"u}r
  Informatik.

\bibitem[DP20c]{Das20ppdp}
Ankush Das and Frank Pfenning.
\newblock {Verified Linear Session-Typed Concurrent Programming}.
\newblock In {\em Proceedings of the 22nd International Symposium on Principles
  and Practice of Declarative Programming}, PPDP '20, New York, NY, USA, 2020.
  Association for Computing Machinery.

\bibitem[DP20d]{Derakhshan20arxiv}
Farzaneh Derakhshan and Frank Pfenning.
\newblock {Circular Proofs in First-Order Linear Logic with Least and Greatest
  Fixed Points}.
\newblock {\em CoRR}, abs/2001.05132, January 2020.

\bibitem[DP22]{DARDHA2022100717}
Ornela Dardha and Jorge~A. Pérez.
\newblock Comparing type systems for deadlock freedom.
\newblock {\em Journal of Logical and Algebraic Methods in Programming},
  124:100717, 2022.

\bibitem[DPP20]{DeYoung2020fscd}
Henry DeYoung, Frank Pfenning, and Klaas Pruiksma.
\newblock {Semi-Axiomatic Sequent Calculus}.
\newblock In Zena~M. Ariola, editor, {\em 5th International Conference on
  Formal Structures for Computation and Deduction (FSCD 2020)}, volume 167 of
  {\em Leibniz International Proceedings in Informatics (LIPIcs)}, pages
  29:1--29:22, Dagstuhl, Germany, 2020. Schloss Dagstuhl--Leibniz-Zentrum
  f{\"u}r Informatik.

\bibitem[DS06]{Deng06}
Yuxin Deng and Davide Sangiorgi.
\newblock Ensuring termination by typability.
\newblock {\em Information and Computation}, 204(7):1045--1082, 2006.

\bibitem[DTK{\etalchar{+}}19]{Dreyer}
Derek Dreyer, Amin Timany, Robbert Krebbers, Lars Birkedal, and Ralf Jung.
\newblock {What Type Soundness Theorem Do You Really Want to Prove?}, October
  2019.

\bibitem[GHP09]{Ghani09}
Neil Ghani, Peter~G. Hancock, and Dirk Pattinson.
\newblock Representations of stream processors using nested fixed points.
\newblock {\em Log. Methods Comput. Sci.}, 5(3), 2009.

\bibitem[GKL14]{Giachino14}
Elena Giachino, Naoki Kobayashi, and Cosimo Laneve.
\newblock Deadlock analysis of unbounded process networks.
\newblock In Paolo Baldan and Daniele Gorla, editors, {\em CONCUR 2014 --
  Concurrency Theory}, pages 63--77, Berlin, Heidelberg, 2014. Springer Berlin
  Heidelberg.

\bibitem[GTL89]{Girard89proofstypes}
Jean-Yves Girard, Paul Taylor, and Yves Lafont.
\newblock {\em {Proofs and Types}}.
\newblock Cambridge University Press, USA, 1989.

\bibitem[Hal85]{Halstead85}
Robert~H. Halstead.
\newblock Multilisp: A language for concurrent symbolic computation.
\newblock {\em ACM Trans. Program. Lang. Syst.}, 7(4):501–538, October 1985.

\bibitem[HLKB21]{Kastberg2021}
Jonas~Kastberg Hinrichsen, Dani\"{e}l Louwrink, Robbert Krebbers, and Jesper
  Bengtson.
\newblock Machine-checked semantic session typing.
\newblock In {\em Proceedings of the 10th ACM SIGPLAN International Conference
  on Certified Programs and Proofs}, CPP 2021, page 178–198, New York, NY,
  USA, 2021. Association for Computing Machinery.

\bibitem[HPS96]{Hughes96popl}
John Hughes, Lars Pareto, and Amr Sabry.
\newblock {Proving the Correctness of Reactive Systems Using Sized Types}.
\newblock In {\em Proceedings of the 23rd ACM SIGPLAN-SIGACT Symposium on
  Principles of Programming Languages}, POPL '96, page 410–423, New York, NY,
  USA, 1996. Association for Computing Machinery.

\bibitem[JJKD17]{Jung2017}
Ralf Jung, Jacques-Henri Jourdan, Robbert Krebbers, and Derek Dreyer.
\newblock Rustbelt: Securing the foundations of the rust programming language.
\newblock {\em Proc. ACM Program. Lang.}, 2(POPL), December 2017.

\bibitem[Kob06]{Kobayashi06}
Naoki Kobayashi.
\newblock A new type system for deadlock-free processes.
\newblock In Christel Baier and Holger Hermanns, editors, {\em CONCUR 2006 --
  Concurrency Theory}, pages 233--247, Berlin, Heidelberg, 2006. Springer
  Berlin Heidelberg.

\bibitem[KPB15]{Krishnaswami15}
Neelakantan~R. Krishnaswami, Pierre Pradic, and Nick Benton.
\newblock Integrating linear and dependent types.
\newblock In {\em Proceedings of the 42nd Annual ACM SIGPLAN-SIGACT Symposium
  on Principles of Programming Languages}, POPL '15, page 17–30, New York,
  NY, USA, 2015. Association for Computing Machinery.

\bibitem[Lev04]{Levy04}
Paul~Blain Levy.
\newblock {\em Call-By-Push-Value: A Functional/Imperative Synthesis (Semantics
  Structures in Computation, V. 2)}.
\newblock Kluwer Academic Publishers, USA, 2004.

\bibitem[LJBA01]{Lee2001}
Chin~Soon Lee, Neil~D. Jones, and Amir~M. Ben-Amram.
\newblock The size-change principle for program termination.
\newblock {\em SIGPLAN Not.}, 36(3):81–92, January 2001.

\bibitem[LM16]{Lindley2016icfp}
Sam Lindley and J.~Garrett Morris.
\newblock Talking bananas: Structural recursion for session types.
\newblock {\em SIGPLAN Not.}, 51(9):434–447, September 2016.

\bibitem[LR19]{Lepigre2019}
Rodolphe Lepigre and Christophe Raffalli.
\newblock Practical subtyping for {C}urry-style languages.
\newblock {\em ACM Trans. Program. Lang. Syst.}, 41(1), February 2019.

\bibitem[Nak00]{Nakano}
Hiroshi Nakano.
\newblock A modality for recursion.
\newblock In {\em Proceedings Fifteenth Annual IEEE Symposium on Logic in
  Computer Science (Cat. No.99CB36332)}, pages 255--266, 2000.

\bibitem[Pad14]{Padovani14}
Luca Padovani.
\newblock Deadlock and lock freedom in the linear $\pi$-calculus.
\newblock In {\em Proceedings of the Joint Meeting of the Twenty-Third EACSL
  Annual Conference on Computer Science Logic (CSL) and the Twenty-Ninth Annual
  ACM/IEEE Symposium on Logic in Computer Science (LICS)}, CSL-LICS '14, New
  York, NY, USA, 2014. Association for Computing Machinery.

\bibitem[PCPT14]{Perez2014}
Jorge~A. Pérez, Luís Caires, Frank Pfenning, and Bernardo Toninho.
\newblock Linear logical relations and observational equivalences for
  session-based concurrency.
\newblock {\em Information and Computation}, 239:254--302, 2014.

\bibitem[Pfe20]{notes15814}
Frank Pfenning.
\newblock {Types and Programming Languages}, 2020.

\bibitem[Plo73]{Plotkin1973}
Gordon Plotkin.
\newblock {\em {Lambda-Definability and Logical Relations}}.
\newblock Edinburgh University, 1973.

\bibitem[PP20]{Pruiksma20arxiv}
Klaas Pruiksma and Frank Pfenning.
\newblock {Back to Futures}.
\newblock {\em CoRR}, abs/2002.04607, February 2020.

\bibitem[PP21]{Pruiksma2021jlamp}
Klaas Pruiksma and Frank Pfenning.
\newblock {A message-passing interpretation of adjoint logic}.
\newblock {\em Journal of Logical and Algebraic Methods in Programming},
  120:100637, 2021.

\bibitem[Sac14]{Sacchini14}
Jorge~Luis Sacchini.
\newblock {Linear Sized Types in the Calculus of Constructions}.
\newblock In Michael Codish and Eijiro Sumii, editors, {\em Functional and
  Logic Programming}, pages 169--185, Cham, 2014. Springer International
  Publishing.

\bibitem[San06]{Sangiorgi06}
Davide Sangiorgi.
\newblock Termination of processes.
\newblock {\em Mathematical Structures in Computer Science}, 16(1):1–39,
  2006.

\bibitem[Sch77]{Schutte}
Kurt Schutte.
\newblock {\em Proof theory (translation from German by J. N. Crossley)}.
\newblock Springer-Verlag Berlin, 1977.

\bibitem[SD03]{SprengerDam2003}
Christoph Sprenger and Mads Dam.
\newblock {On the Structure of Inductive Reasoning: Circular and Tree-Shaped
  Proofs in the µ-Calculus}.
\newblock In {\em Proceedings of the 6th International Conference on
  Foundations of Software Science and Computation Structures and Joint European
  Conference on Theory and Practice of Software}, FOSSACS'03/ETAPS'03, page
  425–440, Berlin, Heidelberg, 2003. Springer-Verlag.

\bibitem[Som21]{Somayyajula2021tlla}
Siva Somayyajula.
\newblock {Towards Unifying (Co)induction and Structural Control}.
\newblock In {\em {5th International Workshop on Trends in Linear Logic and
  Applications (TLLA 2021)}}, Rome (virtual), Italy, June 2021.

\bibitem[SPTDC16]{Severi16}
Paula Severi, Luca Padovani, Emilio Tuosto, and Mariangiola 
  Dezani-Ciancaglini.
\newblock {On Sessions and Infinite Data}.
\newblock In Alberto Lluch~Lafuente and Jos{\'e} Proen{\c{c}}a, editors, {\em
  Coordination Models and Languages}, pages 245--261, Cham, 2016. Springer
  International Publishing.

\bibitem[TB20]{TellezBrotherston19}
Gadi Tellez and James Brotherston.
\newblock Automatically verifying temporal properties of pointer programs with
  cyclic proof.
\newblock {\em Journal of Automated Reasoning}, 64(3):555--578, 2020.

\bibitem[TCP13]{Toninho13}
Bernardo Toninho, Luis Caires, and Frank Pfenning.
\newblock Higher-order processes, functions, and sessions: A monadic
  integration.
\newblock In Matthias Felleisen and Philippa Gardner, editors, {\em Programming
  Languages and Systems}, pages 350--369, Berlin, Heidelberg, 2013. Springer
  Berlin Heidelberg.

\bibitem[TG03]{Thiemann2003}
Ren{\'e} Thiemann and J{\"u}rgen Giesl.
\newblock Size-change termination for term rewriting.
\newblock In Robert Nieuwenhuis, editor, {\em Rewriting Techniques and
  Applications}, pages 264--278, Berlin, Heidelberg, 2003. Springer Berlin
  Heidelberg.

\bibitem[{The}21]{Coq}
{The Coq Development Team}.
\newblock {The Coq Proof Assistant}, January 2021.

\bibitem[Vez15]{Vezzosi2015types}
Andrea Vezzosi.
\newblock {Total (Co)Programming with Guarded Recursion}.
\newblock In Tarmo Uustalu, editor, {\em 21st International Conference on Types
  for Proofs and Programs (TYPES 2015)}, pages 77--78, Tallinn, Estonia, 2015.
  Institute of Cybernetics at Tallinn University of Technology.

\bibitem[Wad12]{Wadler12jfp}
Philip Wadler.
\newblock {Propositions as sessions}.
\newblock In Peter Thiemann and Robby~Bruce Findler, editors, {\em {ACM}
  {SIGPLAN} International Conference on Functional Programming, ICFP'12,
  Copenhagen, Denmark, September 9-15, 2012}, pages 273--286. {ACM}, 2012.

\bibitem[Xi01]{Xi2001lics}
Hongwei Xi.
\newblock Dependent types for program termination verification.
\newblock In {\em Proceedings 16th Annual IEEE Symposium on Logic in Computer
  Science}, pages 231--242, 2001.

\bibitem[XP99]{Xi99popl}
Hongwei Xi and Frank Pfenning.
\newblock {Dependent Types in Practical Programming}.
\newblock In A.~Aiken, editor, {\em Conference Record of the 26th Symposium on
  Principles of Programming Languages (POPL'99)}, pages 214--227. ACM Press,
  January 1999.

\bibitem[YBH04]{YOSHIDA2004145}
Nobuko Yoshida, Martin Berger, and Kohei Honda.
\newblock Strong normalisation in the $\pi$-calculus.
\newblock {\em Information and Computation}, 191(2):145--202, 2004.

\bibitem[Yos96]{Yoshida96}
Nobuko Yoshida.
\newblock Graph types for monadic mobile processes.
\newblock In V.~Chandru and V.~Vinay, editors, {\em Foundations of Software
  Technology and Theoretical Computer Science}, pages 371--386, Berlin,
  Heidelberg, 1996. Springer Berlin Heidelberg.

\end{thebibliography}
